\documentclass[letterpaper,twocolumn,10pt]{article}
\usepackage{usenix2019_v3}

\usepackage{tikz}
\usepackage{amsmath}
\usepackage{subfigure}
\usepackage{makecell}
\usepackage{mathtools}

\usepackage{filecontents}

\graphicspath{{figures/}}

\newcommand{\note}[1]{{\bf{*** #1 ***}}}
\newcommand{\comment}[1]{{}}
\newcommand{\betree}[1]{{B$^{\epsilon}$--tree}}
\newcommand{\betrees}[1]{{\betree{}s}}

\newcommand{\remove}[1]{}
\newcommand{\ignore}[1]{}
\begin{document}
%-------------------------------------------------------------------------------

%don't want date printed
\date{}

\title{\Large \bf VAT: Asymptotic Cost Analysis for Multi-level Key-Value Stores}

%for single author (just remove % characters)
%\author{
%{\rm Your N.\ Here}\\
%Your Institution
%\and
%{\rm Second Name}\\
%Second Institution
% copy the following lines to add more authors
% \and
% {\rm Name}\\
%Name Institution
%} % end author

\author{
{\rm Nikos Batsaras, Giorgos Saloustros, Anastasios Papagiannis$^1$, 
     Panagiota Fatourou$^1$, and Angelos Bilas$^1$}\\
Institute of Computer Science (ICS),
Foundation for Research and Technology -- Hellas (FORTH), Greece\\
100 N. Plastira Av., Vassilika Vouton, Heraklion, GR-70013, Greece \\
Email: \{nikbats, gesalous, apapag, faturu, bilas\}@ics.forth.gr\\
\\
% copy the following lines to add more authors
% \and
% {\rm Name}\\
%Name Institution
} % end author

\maketitle\footnotetext[1]{Also with the Department of Computer Science,
  University of Crete, Greece}

\maketitle
%-------------------------------------------------------------------------------
\begin{abstract}

Over the past years, there has been an increasing number of
key-value (KV) store designs, each optimizing for a different
set of requirements. Furthermore, with the advancements of
storage technology the design space of KV stores has become
even more complex. More recent KV-store designs target fast
storage devices, such as SSDs and NVM. Most of these designs
aim to reduce amplification during data re-organization by
taking advantage of device characteristics.  However, until
today most analysis of KV-store designs is experimental and
limited to specific design points. This makes it difficult to
compare tradeoffs across different designs, find optimal
configurations and guide future KV-store design.

In this paper, we introduce the Variable
Amplification--Throughput analysis (VAT) to calculate
insert-path amplification and its impact on multi-level
KV-store performance. We use VAT to express the behavior of
several existing design points and to explore tradeoffs that
are not possible or easy to measure experimentally. VAT
indicates that by inserting randomness in the insert-path, KV
stores can reduce amplification by more than 10x for fast
storage devices. Techniques, such as key-value separation and
tiering compaction, reduce amplification by 10x and 5x,
respectively. Additionally, VAT predicts that the advancements
in device technology towards NVM, reduces the benefits from
both using key-value separation and tiering.

\end{abstract}

\section{Introduction}
\label{sec:introduction}

Persistent key value (KV) stores~\cite{decandia2007dynamo,
hbase,rocksdb, leveldb} are a central component for many analytics
processing frameworks and data serving systems. These systems are
considered as write-intensive because they typically exhibit bursty
inserts with large variations in the size and type of data
items~\cite{chen2012interactive, sears2012blsm}. Consequently, over
the last few years, KV stores have evolved to support many different
applications and workloads~\cite{kvdiver}. There has been a number of
new techniques that either optimize for different uses, e.g. write vs.
read vs. scan or optimize certain aspects of system operation. As a
result, this has increased the complexity of the KV-store design space
to a point that it is unclear how each technique affects performance.
Better understanding of KV-store design tradeoffs has the potential to
improve both application performance and data serving infrastructure
efficiency.

KV stores typically use at their core the write-optimized
LSM-Tree~\cite{lsm} to handle bursty inserts and amortize write I/O
costs. LSM-Tree~\cite{lsm} organizes data in multiple levels of
increasing size (the size ratio of successive levels is known as
growth factor). Each data item travels through levels until it reaches
the last level. Data items generally move in batches from level to
level with a merge/sort operation (compaction) that reorganizes data
across levels. Each level is further physically organized into
segments called \emph{sorted string tables (SSTables or SSTs)}. Each
SST stores a \emph{non-overlapping and sorted} subset of the key
space.

Traditionally, such multi-level KV stores target Hard Disk Drives
(HDDs) as the storage medium, because HDDs exhibit lower cost per bit
of stored information. However, in HDDs, random I/O requests have a
substantial negative effect on device throughput. For this reason,
multi-level KV stores use large SSTs to always generate large I/O
requests. Large SSTs have two important benefits: First, due to SST's
large size (in the order of MB), KV stores issue only large I/Os to
the storage devices resulting in optimal HDD throughput. Second, they
require a small amount of metadata for book-keeping due to their
sorted and non-overlapping nature. SST metadata fit in memory and are
only modified during compactions, thus they do not generate random
I/Os in the common path. Therefore, multi-level KV stores are
guaranteed to perform only sequential I/O to devices.

On the other hand, a significant drawback of the multi-level design is
its high I/O amplification: The merge operation across levels results
in reading and writing data many more times than the size of the data
itself, resulting in traffic of up to several multiples of 10x
compared to the dataset size~\cite{wisckey}. Although amplification is
so high, it still is the right tradeoff for HDDs: Under small, random
I/O requests, HDD performance degrades by more than two orders of
magnitude, from 100 MB/s to 100s KB/s.

\begin{figure}
\centering
\includegraphics[width=0.24\textwidth]{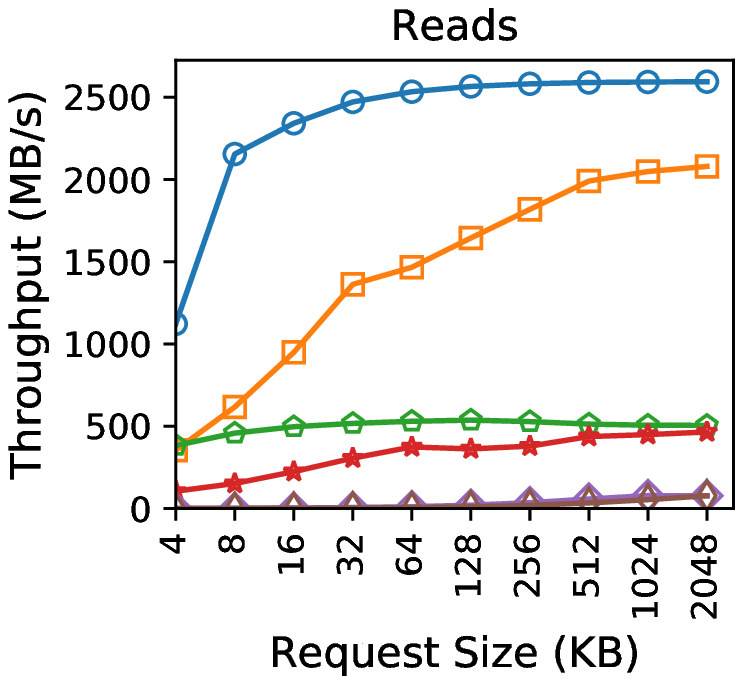}~
\includegraphics[width=0.24\textwidth]{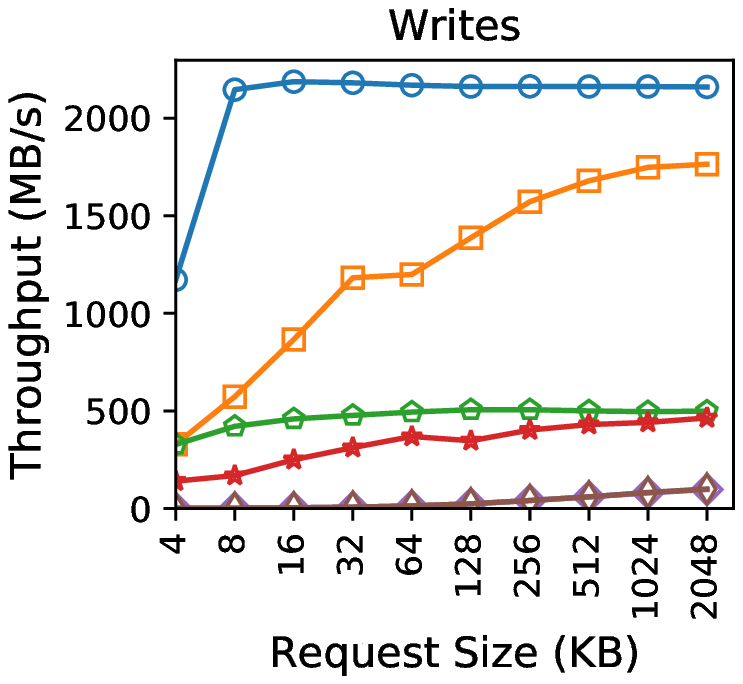}
\newline
\includegraphics[width=0.48\textwidth]{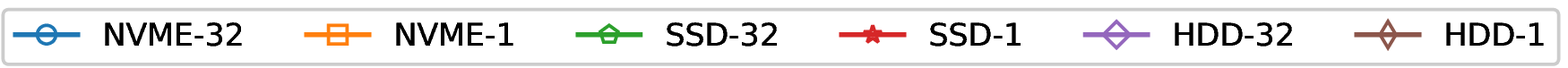}
\caption{FIO~\cite{fiobenchmark} throughput vs. request size, using
  iodepth 1 and 32, for three different device technologies: HDD
  (Western Digital Black Caviar 4 TB), SSD (Samsung 850 Pro 256 GB),
  and NVMe (Intel Optane P4800X 375 GB).}
\label{fig:fio}
\end{figure}

With the emergence of fast storage devices, such as NAND-Flash solid
state drives (SSDs) and non-volatile memory devices (NVMe), the design
space of KV stores has grown further. In modern devices, the device
behavior is radically different under small, random I/Os: At
relatively high concurrency, most of these devices achieve a
significant percentage of their maximum throughput
(Figure~\ref{fig:fio}). At the same time, introducing some level of
random I/Os can reduce amplification. Previous
work~\cite{wisckey,tucana,kreon,hashkv} has used this property of
graceful device throughput degradation with random I/O to demonstrate
the benefits of reducing I/O amplification and introducing various
techniques, such as key-value separation and small SSTs with B+-tree
indexing.  These systems essentially draw a different tradeoff between
\textit{device throughput} and \textit{amplification}. Therefore,
modern storage devices dictate different designs for KV stores that
draw a different balance between amplification and throughput to
achieve higher performance in the \emph{insert-path,} further
increasing the complexity of the design space. Such designs can reduce
the amount of data reorganization and therefore, they have the
potential to both increase device efficiency and reduce CPU
utilization.

Although these efforts derive from the original LSM design~\cite{lsm},
they cannot be described by the LSM cost analysis, since it assumes
that the system performs only large and practically fully sequential
I/Os at the cost of performing a full read/write of two successive
levels during each merge operation. Currently, there is no analysis
that captures the tradeoffs between device throughput and
amplification and reflects the cost for all these designs. The lack of
such an analysis makes it more difficult to reason for tradeoffs
across techniques and thus, navigate the design space and identify
improved design points for new sets of requirements.

In this paper, we present VAT, a cost analysis for the insert-path
that describes different techniques, such as leveling and
tiering~\cite{stepmerge} compaction and performing key-value
separation using value logs. VAT also captures the tradeoff of
variable (as opposed to maximum) amplification vs. variable (as
opposed to maximum) device throughput. We use VAT to derive optimal
values for level growth factor, quantify the benefits of different
design points, analyze tradeoffs, make projections and guide KV stores
towards optimal design configurations.

Our VAT analysis, similar to the original LSM-Tree cost
analysis~\cite{lsm}, describes the operation of a multi-level system
as a series of data transfers. Unlike the original analysis though,
VAT introduces additional parameters for modeling different techniques
as well as variable amplification and achieved device throughput. We
use the VAT analysis to derive and solve a minimization problem that
can quantify various aspects of KV-store designs, including the use of
fast storage devices. In our analysis, we determine optimal values for
the number of levels ($l$) and growth factor between levels ($f$), we
examine differences across designs, and explore trends as device
technology evolves. We find that by inserting randomness in the
design, amplification drops by more than 10x, using a log can reduce
amplification by 10x for small key-value size ratios of up to 1\% and
using tiering~\cite{stepmerge} instead of leveling decreases
amplification by 5x, at the cost of read and scan operations.

Our main contributions are:
\begin{itemize}
\item 
We present VAT, an asymptotic cost analysis that captures data
amplification in the insert-path for a wide collection of KV-store
designs, including designs dictated by modern device technology. VAT
can be extended to also capture additional design points in the
future.
\item 
We perform a comprehensive experimental analysis to show that VAT
captures the behavior of four well-known KV-store systems,
RocksDB~\cite{rocksdb}, Kreon~\cite{kreon}, BlobDB~\cite{blobdb},
PebblesDB~\cite{pebblesdb}, and that it accurately predicts e.g. the
optimal level configuration for each design.  Thus, VAT is a useful
tool to understand tradeoffs between existing systems and
configurations.
\item 
VAT allows us to better understand: (a) the effects of device
technology and randomness on I/O amplification and data reorganization
cost, (b) optimal values for important parameters of existing KV
stores and asymptotic trends, (c) tradeoffs between different design
techniques, and (d) I/O and space amplification tradeoffs.
\end{itemize}

The rest of this paper is organized as follows:
Section~\ref{sec:background} briefly discusses necessary background,
Section~\ref{sec:vat} presents VAT and the thought process behind the
analysis, Section~\ref{sec:meth} presents our experimental
methodology, Sections~\ref{sec:eval} presents our experimental
results, and Section~\ref{sec:proj} discusses tradeoff analysis and
VAT projections. Finally, Section~\ref{sec:related} reviews related
work and Section~\ref{sec:conclusions} concludes the paper. We also
present supplementary detailed derivations for VAT and LSM-Tree
analysis in appendices~\ref{app:vat} and~\ref{app:lsm}.

\section{Background}
\label{sec:background}

A Log-Structured Merge (LSM) tree~\cite{lsm} is a multi-level data
structure that optimizes for bursty inserts. The first level ($L_0$)
is memory resident, whereas the rest of the levels are on the device.
We assume that the LSM structure has $l$ levels and we denote by
$S_i$ the size of level $L_i$ ($0 \leq i \leq l$). In the original LSM
paper~\cite{lsm} the levels have exponentially increasing sizes: For
each $1 < i \leq l$, $L_i$ contains $f$ times more data than  the
previous level $L_{i-1}$; where $f$ is the {\em growth factor}.  Each
level consists of a set of {\em sorted string tables} (SSTs)
containing (key,value) pairs.

During an insert operation, LSM stores the key value pair is in the
$L_0$ memory component. Periodically, the data of a level are flushed
to the next (lower) level, to free up space in upper levels. This
process is called {\em compaction} which produces excess read and
write I/O traffic named {\em amplification}.

There are different ways to organize data across levels. In the {\em
leveling} approach, each level organizes its key value pairs in
non-overlapping SSTs. SSTs can be small and sequentially placed on
the device~\cite{sears2012blsm}, large(tenths of MB) and placed
randomly~\cite{rocksdb}, or small and random~\cite{kreon}. In leveling,
a compaction merge-sorts SSTs from the upper level with overlapping
SSTs from the lower level. Because of sorting, this incurs high I/O
amplification and CPU overhead. In the {\em tiering}
approach~\cite{stepmerge}, there are overlaps in the key ranges of
different SSTs within a level. When a compaction is triggered on a
level, the SSTs residing on the level are sorted and moved to the next
level, without performing any merging. Thus, tiering incurs
significantly less I/O amplification but lower read performance than
leveling due to overlapping of SSTs per level.

A set of systems~\cite{atlas,wisckey,tucana,kreon,hashkv} instead of
storing values with keys in each level, they use KV separation. This
technique appends values in a value log and only re-organize the keys
(and pointers) in the multilevel structure. Note that the second
technique incurs lower amplification than the first.

\section{VAT: Variable Amplification-Throughput Analysis}
\label{sec:vat}

In this section, we present the VAT asymptotic analysis to model
different techniques of the multi-level KV-store design space and
capture the variable amplification and achieved throughput tradeoff in
fast storage devices. Specifically, VAT describes four categories of
designs: leveling and tiering, with and without key-value separation
(value log). 

\subsection{Modeling I/O traffic}
\label{sec:traffic}

The VAT basic equation, which we use to derive all subsequent
relations, captures two major costs related to amplification in the
insert-path: First, the cost that corresponds to the amount of excess
I/O traffic generated during merge operations of two adjacent levels,
the lower (larger) level and the upper (smaller) level. The basic
equation measures this cost under the assumption that during a merge
operation, the lower level is fully read and written. It is important
to notice, that we refine this assumption in subsequent VAT equations.
As a result, assuming the lower level is $f$ times larger than the
upper level, the system reads and writes an excess of $f$ times more
bytes, compared to the upper level. Second, the cost of data
reorganization across levels: In a system with $l$ levels, each data
item moves through all levels resulting in $l$ times excess traffic.
We refer to these quantities of excess traffic as
\textbf{\textit{merge amplification}} and \textbf{\textit{level
amplification}}, respectively.

If $S_0$ is the size of the in-memory first level and $S_l$ is the
size of the last level, then we can assume that the entire
workload/dataset fits in (is equal to) $S_l$ and that all data will
eventually move to the last level $S_l$. Then, $S_l/S_i$ is the total
number of merge operations from $L_i$ to $L_{i+1}$, until all data
reach $L_l$.

The basic equation measures the amount of I/O traffic $D$ produced
until all $S_l$ data reach $L_l$:
\vspace*{-.3cm}
\begin{eqnarray}
D &= & \frac{S_l}{S_0}(S_0)  + 2\sum_{j=1}^{S_l/S_0}((j-1)\bmod{f})\cdot S_0\nonumber\\
& + & \frac{S_l}{S_1}(2S_1)  + 2\sum_{j=1}^{S_l/S_1}((j-1)\bmod{f})\cdot S_1\nonumber\\
& + & \ldots\nonumber\\
& + & \frac{S_l}{S_{l-1}}(2S_{l-1}) + 2\sum_{j=1}^{S_l/S_{l-1}}((j-1)\bmod{f})\cdot S_{l-1}
\label{eq:vatamp}
\end{eqnarray}

\vspace*{-.2cm}
In equation~(\ref{eq:vatamp}), each sub-expression (row) captures the
amplification of merging between two consecutive levels. For each such
merge operation, there are two terms. The first term represents the
data of the upper (smaller) level that have to be read and written
during the merge operation, while the second term represents the data
of the lower (larger) level. 

For each level $L_i$, $0 \leq i \leq l-1,$ each time one of the $S_l /
S_i$ merge operations occurs, all data stored in $L_i$ are read and
written, thus causing I/O traffic of size $2S_i$. This explains the
first term that appears in each row. Also note that, in the first
sub-expression for $L_0$ that resides in memory, the factor of $2$ is
missing in the first term, indicating that we do not perform I/O to
read data that are already in memory.

The second term captures the total amount of data that are read and
written from $L_{i+1}$ in order to merge the overlapping ranges of
$L_i$ and $L_{i+1}$.  It uses  the $\sum$ operator to express the fact
that the merge operation happens multiple times as data flow through
the system. It also uses the $\bmod$ operator to capture the fact that
the size of the lower (larger) level grows incrementally up to $f$: in
the first merge operation the lower level has no data (i.e., $j-1=0$);
in the next merge, the lower level contains data equal to 1x the upper
level; in each subsequent merge operation it contains data 2x, 3x,
etc. of the data in the upper level. These data need to be read and
written during merge, hence the factor of $2$ before the sum.

\subsection{Modeling Variable Amplification and Device Throughput}
Equation~(\ref{eq:vatamp}) assumes that the merge operation of two
adjacent levels incurs the maximum possible amplification by reading
and writing them in full. However, currently there are KV-store
designs~\cite{tucana,kreon,wisckey,hashkv,lsmtrie,shetty2013building}
targeting fast storage devices (SSD, NVMe), which draw a different
balance between amplification and device throughput. In particular,
they reduce amplification by taking advantage of the property of fast
storage devices to perform close to sequential throughput under random
I/O pattern. VAT effectively captures this relationship between lower
amplification at increased randomness and reflects the cost of various
design decisions. It does so by introducing parameters $a$, $r$. The
$a$ parameter models the impact of the SST size, the data organization
technique, and the input, to amplification. In particular, $a$
expresses the percentage of the size of the lower level which is read
and written during compaction. The $r$ parameter expresses the
achieved device throughput and depends on the SST size and the degree
of concurrency on the device. Both $a,r$ are in the range [0,1] (and
$r\neq0$). We call $a$ the {\em merge amplification parameter} and $r$
the {\em achieved throughput parameter}. Below we discuss how the SST
size, data organization, and input affect $a$ and $r$.

\vspace*{.1cm}
\noindent
\underline{{\it SST Size:}}
The use of small SSTs leads to a fine grained partitioned level with
more SSTs per level. This allows to use techniques that reduce merge
amplification significantly. For instance, reducing the SST size for
cases where the input distribution is zipf can lead to reading a
smaller percentage of the lower level at each compaction.  This is
because during a compaction, there is higher chance to find a hole in
the key space of the lower level and thus choose to merge the SST of
the upper level that fills that hole, therefore making the merge
process cheaper in terms of I/O traffic produced.

\vspace*{.1cm}
\noindent
\underline{{\it Data organization:}} 
Leveling compaction~\cite{lsm} merges two levels keeping each level
physically sorted in large chunks on the device. Therefore, the value
of the merge amplification parameter $a$ is 1. Previous works have
proposed various
techniques~\cite{kreon,shetty2013building,lsmtrie,pebblesdb,sifrdb} to
lower the value of $a$. For instance, the use of \emph{compaction
priorities} in RocksDB~\cite{rocksdb} tries to change the order in
which SSTs are merged to allow for merging, more frequently, SSTs to
lower levels that are less full. This effectively reduces $a.$
Similarly, the use of an index~\cite{kreon} allows for smaller SSTs
that are not necessarily contiguous on the device, therefore providing
opportunities to reduce the amount of data amplification during merge
operations. On the other hand, in tiering, merging happens only in the
level that triggered the compaction, thus merge amplification $a$ is
$0.$

\vspace*{.1cm}
\noindent
\underline{{\it Input:}}
The input distribution, the ordering of keys and the percentage of
updates, affect the overlap of keys in each pair of levels that will
be merged. For instance, a uniform distribution where each (large) SST
contains keys from the entire key space will result in maximum merge
amplification ($a=1$), whereas a sorted input sequence of keys will
result in $a=0.$

\vspace*{.2cm}
\noindent
{\bf Determining $a$ and $r$ experimentally:}
Parameter $r$ depends mainly on the SST size. As a result, one can
estimate the value of $r$ for a given KV-store design by using a
micro-benchmark (e.g. FIO~\cite{fiobenchmark}) which simulates SST
size and random I/O access pattern. Parameter $r$ value is the ratio
of the measured throughput over the sequential. Note that $r$ also
depends on the degree of concurrency: higher degrees of concurrency
results in increased values for $r$. However, since most current
systems achieve a high degree of concurrency, its effect on $r$ is
more or less the same in all systems. As a result, we determine $r$ as
a function only of the SST size and the device type. We calculate the
value of $r$ as the percentage of the achieved throughput when random
I/O operations of size equal to the target SST size are performed on
the device (under high concurrency) over its sequential. Previous
work~\cite{tucana,kreon} has reported that KV stores are able to
generate I/O queue depths of around 30, so we determine $r$ using a
similar value. These values also agree with the experiments we
performed with different KV stores in the context of this work.

Determining a realistic value for parameter $a$ is more involved.
However, we observe that although the value of parameter $a$ depends
on the KV-store design, it does not depend on the device technology:
Although adaptive KV-store designs might be possible in the future,
currently, KV stores do not adjust their main data structures and
operations when using different devices. Thus, to determine the value
of $a,$ we measure the amount of excess bytes during compaction with
the lower levels by experimentally performing the following
measurement. For each merge operation, we calculate:

\vspace*{-.2cm}
\begin{equation*}
	a =
	\frac{MSST_{L}}{MSST_{U}(TSST_{L}/TSST_{U})}\, ,
\end{equation*}

\vspace*{-.1cm}
\noindent
where $MSST_L$ and $MSST_U$ are the numbers of SSTs of the lower and
the upper level, respectively, that participate in compaction, and
$TSST_L,$ $TSST_U$ are the numbers of total SSTs stored in the lower
and upper level, respectively, at the time of merging. We then
calculate the mean of all such values (over all the executed
compactions) to get the estimated value for $a.$
Table~\ref{tab:atable} presents the result of the above measurement
for the following systems: RocksDB~\cite{rocksdb}, Kreon~\cite{kreon},
BlobDB~\cite{blobdb}, and PebblesDB~\cite{pebblesdb}. Typically, each
KV store uses a specific growth factor $f,$ at around 10. We choose
the growth factor $f=8$ since it is close to 10 and it is the growth
factor that makes the capacity of the last level equal to the workload
size. We discuss these values further in our results.

\begin{table}[t]
\centering
\begin{tabular}{|c|c|c|c|c|}
\hline
$f$ & RocksDB & Kreon & BlobDB & PebblesDB \\
\hline
8  & 0.68 & 0.25 & 0.8 & 0 \\
\hline
\end{tabular}
\caption{Merge amplification $a$ for growth factor $f=8.$}
\label{tab:atable}
\end{table}

\subsection{The VAT Cost Analysis}

In this section, we present the VAT cost analysis. We first provide
the basic VAT analysis and then we present similar equations for
different designs, like key-value separation with a value log and
tiering. 

\paragraph{Basic VAT Analysis:}
VAT calculates the time $T$ to write the data of size $S_l$ (that fit
in the last level) and the optimal time $T_{opt}$ as follows:

\vspace*{-.3cm}
\begin{equation} 
	\label{eq:T and Topt}
	T = \frac{D}{r \cdot R_{opt}} \hspace*{.5cm} \mbox{    and     } \hspace*{.5cm} 
	T_{opt}=\frac{S_l}{R_{opt}} 
\end{equation}

\vspace*{-.2cm}
$R_{opt}$ is the optimal device throughput and $r\cdot R_{opt}$ is the
achieved device throughput (recall that $r$ is in the range (0,1]).
The optimal time to write the data can be expressed as the ratio of
the minimum amount of data $S_l$ to be written with the maximum
possible throughput $R_{opt}$, as would e.g. be the case for appending
all data in a log file.

Now, we derive the base VAT analysis. By inserting $a$ in
Equation~(\ref{eq:vatamp}) we get:
\vspace*{-.3cm}
\begin{eqnarray}
D &= &\frac{S_l}{S_0}(S_0)  + 2a\sum_{j=1}^{S_l/S_0}((j-1)\bmod{f})\cdot S_0\nonumber\\
  &+ &\ldots\nonumber\\
  &+ &\frac{S_l}{S_{l-1}}(2S_{l-1}) + 2a\sum_{j=1}^{S_l/S_{l-1}}((j-1)\bmod{f})\cdot S_{l-1}\label{eq:vatampa}
\end{eqnarray}

\vspace*{-.1cm}
Note that $S_l/S_{l-i} = f^i$ (assuming $f$ is constant). Using this,
we perform arithmetic transformations to analyze the sum (see
Appendix~\ref{app:vatdata}):

\vspace*{-.3cm}
\begin{equation}
\label{eq:vatdata}
	D = S_l(2l -1 -al + afl)
\end{equation}

Ideally, we would like a KV-store design to minimize the quantity
$\frac{T}{T_{opt}}$ and achieve a time as close to optimal as
possible. For this reason, VAT focuses on the following minimization
problem: $\displaystyle\min_{\substack{0\le a\le 1\\0<r\le1}}
\frac{T}{T_{opt}}$. Using Equation~(\ref{eq:vatdata}), we get:

\begin{equation}
  \frac{T}{T_{opt}} = \frac{2l -1 - al + afl}{r}\label{eq:vat}
\end{equation}
Considering $a$ and $r$ are fixed values for a given design,
Equation~(\ref{eq:vat}) expresses $T/T_{opt}$ as a function of $l$ and
$f$. So, by studying the minimization problem, we gain insight in the
tradeoff between the number of levels and the growth factor.

\paragraph{VAT for key-value separation:}
We now apply the VAT analysis (presented above) to express key-value
separation using a value log. Denote by $K_i$ and $V_i$ the total size
of keys and values, respectively, of each level $L_i.$ Note that each
SST now stores only keys and thus its level is equal to $K_i.$
However, in our equations below, we let $S_i = K_i + V_i.$ The value
log contains all the key-value pairs stored in the system, so its size
is $S_l.$ Following a similar approach as above, we express the total
I/O traffic as:
	
\vspace*{-.3cm}
\begin{eqnarray}
D &= & \frac{K_l}{K_0}(K_0)  + 2a\sum_{j=1}^{K_l/K_0}((j-1)\bmod{f})\cdot K_0\nonumber\\
	& + & \ldots\nonumber\\
&+ & \frac{K_l}{K_{l-1}}(2K_{l-1}) + 2a\sum_{j=1}^{K_l/K_{l-1}}((j-1)\bmod{f})\cdot K_{l-1}\nonumber\\
& + & S_l \label{eq:vatlogamp}
\end{eqnarray}

The last term $S_l$ in Equation~(\ref{eq:vatlogamp}) accounts for
appending the entire dataset in the value log. Let $p = K_l/V_l$.
Then, $p$ is typically a small constant ($0 < p < 1$). Using
Equations~(\ref{eq:T and Topt}), we get (see Appendix~\ref{app:vatlog}
for the derivations):

\vspace*{-.5cm}
\begin{eqnarray}
D &= & K_l\Big(2l -1 - al + afl\Big) + S_l\label{eq:vatlogdata}\\
\frac{T}{T_{opt}} &= & \frac{p(2l -1 - al + afl) + p + 1}{r\cdot (p+1)}\label{eq:vatlog}
\end{eqnarray}

KV stores are used to support diverse workloads~\cite{kvdiver}, where
key and value sizes may differ within a wide range: in typical
workloads, keys are a few tens of bytes, whereas values vary from
similar sizes to a few KB of data (thus resulting in values of $p$
much smaller than $1$).  Therefore, in our evaluation we examine
various data points where key to value size ratio spans the range from
1 to 0.01. 

Note that for small values of $p$, e.g. close to 0.01, the ratio in
Equation~(\ref{eq:vatlog}) is much smaller (the numerator is smaller
than the denominator) than that in Equation~(\ref{eq:vat}). This way,
VAT shows that using key-value separation with a value log has a
significant benefit in terms of incurred amplification.

\paragraph{VAT for tiering:}
In tiering compaction, excess traffic during merge operations includes
only reading and writing the data in $L_i$ (and not $L_{i+1}$ as in
leveling). Therefore, $a =0$. By setting $a=0$ in
Equation~(\ref{eq:vat}), we get the equations for tiering:

\vspace*{-.2cm}
\begin{eqnarray}
	\frac{T}{T_{opt}} &= & \frac{2l-1}{r} = \frac{2\log_f C-1}{r}\, ,  \label{eq:vattier}
\end{eqnarray}

\vspace*{-.2cm}
\noindent
where $C = S_l / S_0 = f^l$ (and therefore $l = \log_f C$).

Equivalently to Equation~(\ref{eq:vatlogamp}), to model the cost of
tiering with key-value separation, we slightly modify the analysis
above to only consider keys ($K_l$) instead of both keys and values
($S_l$).

\vspace*{-.3cm}
\begin{align}
	\frac{T}{T_{opt}} = \frac{p(2l-1)+p+1}{r\cdot(p+1)} =
	\frac{p(2log_fC-1) + p + 1}{r\cdot(p+1)}
	\label{eq:vattierlog}
\end{align}

\vspace*{-.2cm}
These equations express the fact that tiering does not depend on the
size of the next level and therefore, on $a,$ which expresses excess
bytes related to the next level. Consequently, tiering cannot benefit
as much as leveling from emerging device technologies in the
insert-path.

\paragraph{VAT Equations:}
In summary, the VAT analysis can describe different designs, in terms
of cost for the insert-path, as shown in Table~\ref{tab:vat}.

\begin{table}[t]
	\centering
	\begin{tabular}{|l|c|c|}
		\hline
		&no log & log \\ \hline
		VAT $T/T_{opt}$ & $\frac{2l-1 - al + afl}{r}$
		& $\frac{p(2l-1 - al + afl) + p + 1}{r\cdot (p+1)}$ \\
		\hline
	\end{tabular}
	\vspace*{-.1cm}
	\caption{VAT equations for the minimization problem $T/T_{opt}$.}
	\label{tab:vat}
	\vspace*{-.3cm}
\end{table}

\section{Experimental methodology}
\label{sec:meth}

In our evaluation, we examine two main aspects of VAT:
\begin{itemize}
\item How accurately VAT can model the behavior of different
  techniques. We use four existing KV-store systems to examine how
  accurately VAT can model different points in the design
  configuration space: RocksDB~\cite{rocksdb}, Kreon~\cite{kreon},
  BlobDB~\cite{blobdb}, and PebblesDB~\cite{pebblesdb}.
\item How VAT can help understand tradeoffs between different design
  points. For this purpose we quantify the benefits of different
  designs, present observations on their asymptotic behavior and make
  projections as device technology improves.
\end{itemize}

The real systems we use in our measurements incur significant
complexity, especially systems such as RocksDB that is used
extensively in real-life applications and support many different modes
of operation. Next, we discuss how we modify or configure each system
for our purposes.

\paragraph{RocksDB:} RocksDB by default performs leveling compaction
with values in-place with keys. However, RocksDB can also operate in
different modes and use several techniques that try to reduce
amplification in a \emph{non-asymptotic} manner. Given that VAT models
asymptotic behavior, we make a number of modifications to RocksDB
configuration and code to disable certain non-asymptotic
optimizations: We modify RocksDB to move all SSTs of intermediate
levels to the last level upon termination, to better approximate
steady state operation with large workloads, similar to what VAT
models. We disable the \emph{Write Ahead Log (WAL)} mechanism, since
we measure the I/O traffic produced solely by compactions. We
configure RocksDB to perform leveling compaction with different growth
factors. We use the default RocksDB configuration for memtables
(2x64MB) and $L_0, L_1$ size (256MB) with a maximum of 4 SSTs in
$L_0$. Therefore, only levels greater or equal to $L_1$ exhibit the
prescribed growth factor with respect to the previous level.
Essentially, value $l$ in the asymptotic analysis of LSM and VAT
corresponds to $l+1$ in RocksDB, since in the cost model, levels $L_0$
and $L_1$ also obey the growth factor $f$.

\paragraph{Kreon:} Kreon uses a value log for key-value separation and
organizes the metadata using a multi-level index. We use a 256MB
$L_0$, which is more than enough to hold the metadata of a 16GB
workload in memory.  Kreon by default stores 200 keys per SST. With a
uniform distribution, merge amplification $a$ is close to 1. However,
by only decreasing the number of keys stored per SST down to 4,
resulting in more SSTs per level, thus a more fine grained partitioned
level, the system achieves lower merge amplification with $a=0.25.$

\paragraph{BlobDB:} BlobDB is a wrapper of RocksDB that employs
key-value separation using a value log. It stores values in blobs (log
files) and keys along with metadata in RocksDB's LSM index.  We use
the same configuration with RocksDB.

\paragraph{PebblesDB:} PebblesDB is built on top of
LevelDB~\cite{leveldb} and features tiering compaction. We perform
similar modifications to better approximate steady state operation.
PebblesDB uses the notion of \emph{guards}. Each guard has a set of
associated SSTs and divides the key space (for that level) into
disjoint units. Guards within a level never have overlapping key
ranges. The growth factor in PebblesDB is the number of SSTs in a
guard that triggers the compaction.

\paragraph{Workload:} In our measurements we use a single
YCSB~\cite{Jinglei2016} thread to produce a workload of 16 million
key-value pairs with a value size of $\approx$ 1 KB (1079 bytes). We
generate uniformly distributed keys over a configurable key universe.
We limit the key universe to only contain 3-byte keys and we generate
all 16 million keys in the full range. To do this, we sort the keys
and then we use a stride equal to the ratio of the key universe range
over the number of keys in each SST, to cover the full key universe in
a uniform manner, for every SST that is generated. We run our
experiments using a single database on a Samsung SSD 850 PRO 256GB.
For each run, we vary the growth factor and let the KV stores spawn
the corresponding levels.

We calculate I/O amplification as follows. We measure I/O traffic
externally to the KV stores, using iostat to ensure we capture all
device traffic. We disable the use of the buffer cache so that all
traffic to the devices is visible to iostat. We measure the read and
write traffic in bytes during each experiment. We calculate the
amplification ratio by dividing I/O traffic with the size of the YCSB
dataset ($(1079+3)\cdot 16M$). For VAT, we set $r=1,$ for RocksDB,
BlobDB and PebblesDB since they use large SSTs. Kreon produces 8KB
requests. We used FIO~\cite{fiobenchmark} and measured the achieved
performance with 8KB requests to be $r=0.91$ on the NVMe device we
used for the experiments. Equation~\ref{eq:tovertopt} shows that
$\frac{T}{T_{opt}}$ can also be expressed as a ratio of bytes:

\vspace*{-.2cm}
\begin{equation}
	\frac{T}{T_{opt}} = \frac{ D/(r\cdot R_{opt}) }{S_l/R_{opt}} 
        = \frac{D}{r\cdot S_l}
\label{eq:tovertopt}
\end{equation}

\section{VAT on several KV-store designs}
\label{sec:eval}

\begin{figure}
\centering
\subfigure[RocksDB]{\includegraphics[width=0.24\textwidth]{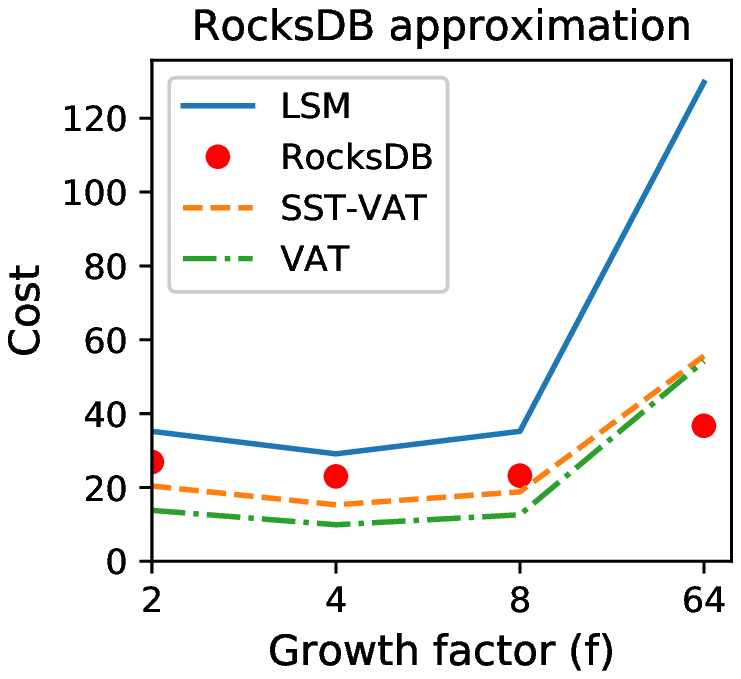}}~
\subfigure[Kreon]{\includegraphics[width=0.24\textwidth]{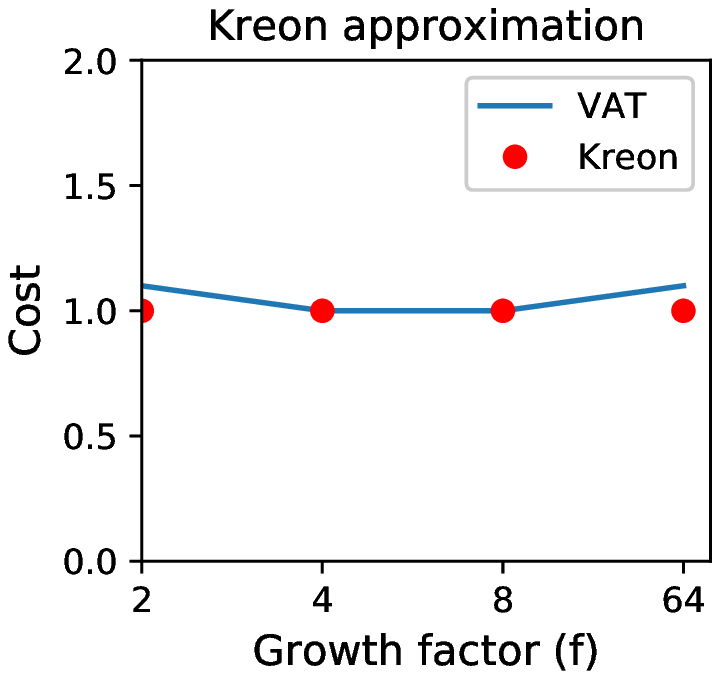}}
\subfigure[BlobDB]{\includegraphics[width=0.24\textwidth]{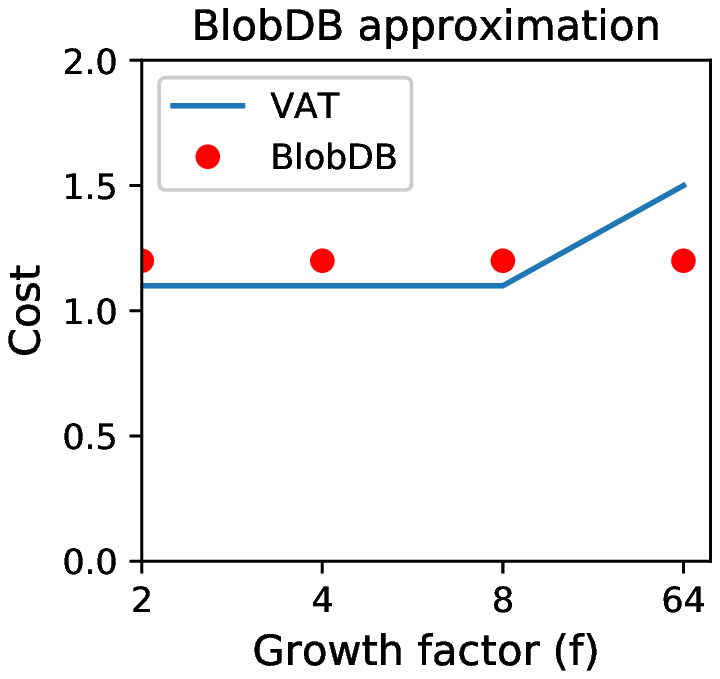}}~
\subfigure[PebblesDB]{\includegraphics[width=0.24\textwidth]{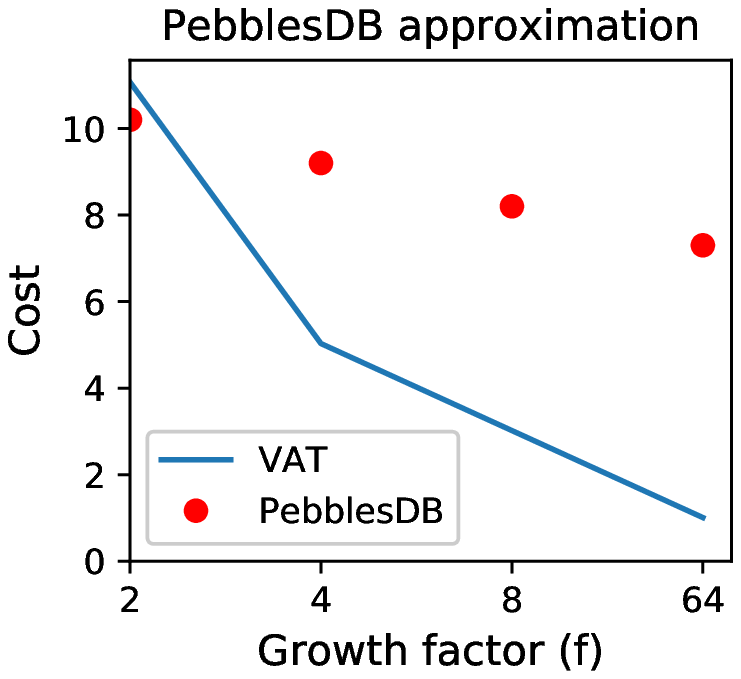}}
\caption{VAT models for different KV-store designs and actual
  measurements.}
\label{fig:VATapprox}
\end{figure}

In this section, we use measurements from the four KV stores in our
experimental setup to show that VAT captures the behavior of different
design points and is able to suggest the optimal level configuration
for each technique. Figure~\ref{fig:VATapprox} summarizes our
experimental results, which we discuss below.

\begin{figure}
\centering
\subfigure[$r=1$,~$C=1000$]{\includegraphics[width=0.24\textwidth]{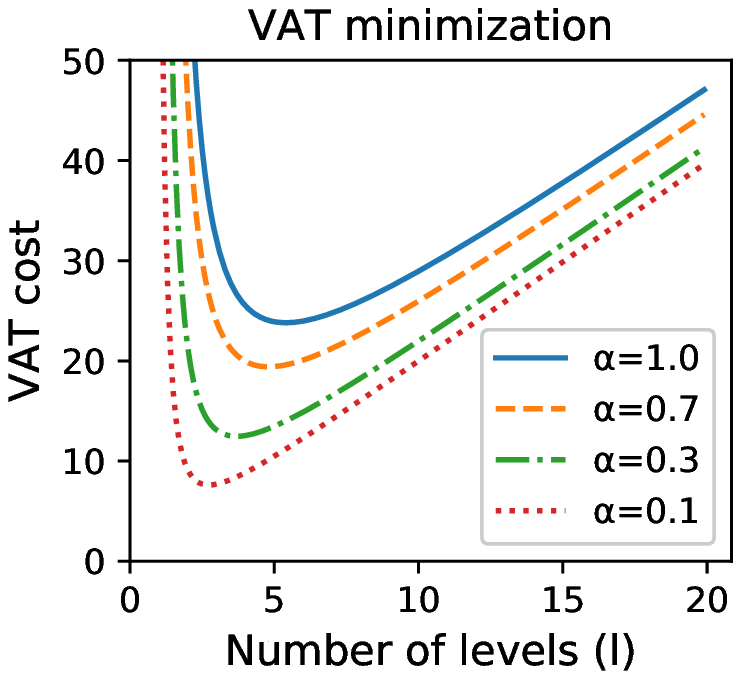}}~
\subfigure[$a=1$,~$C=1000$]{\includegraphics[width=0.24\textwidth]{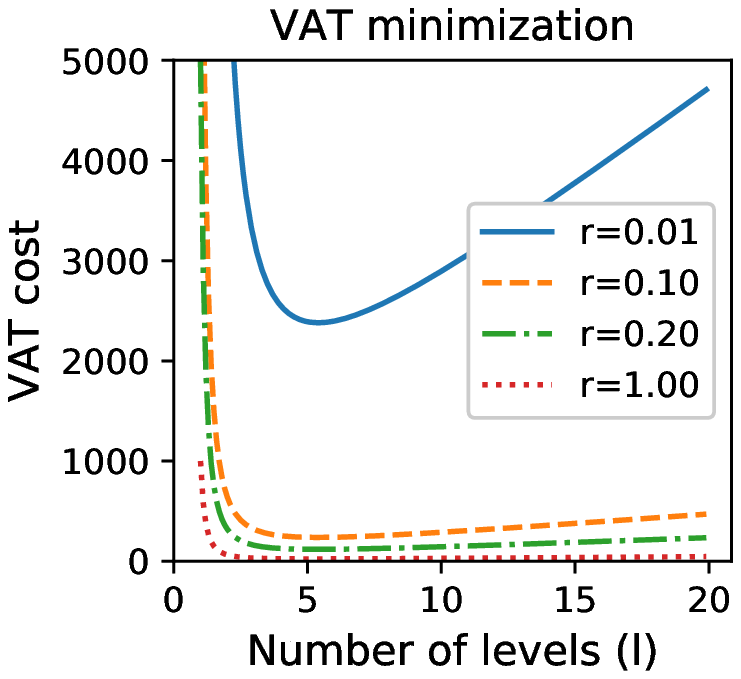}}
\caption{
(a) VAT cost ratio ($T/T_{opt}$) for optimal throughput ($r=1$)
    and different values of merge amplification $a.$
(b) VAT cost ratio at maximum merge amplification ($a=1$)
    and different values of achieved device throughput $r.$
}
\label{fig:VATbase}
\end{figure}

Figure~\ref{fig:VATapprox}(a) shows the cost ratio $\frac{T}{T_{opt}}$
with an increasing number of levels for leveling without a value log
as calculated with VAT and as measured with RocksDB. We also include
the cost as calculated by LSM-Tree analysis. We see that VAT is close
to the actual measurements. 

In addition, RocksDB performs compactions between two levels $L_i$ and
$L_{i+1}$ on a per-SST basis. Therefore, a compaction in RocksDB does
not necessarily merge all SSTs of $L_i$ to $L_{i+1}.$ This results in
all levels being continuously almost full. VAT expresses this behavior
as follows (where $B$ is equal to the SST size):

\vspace*{-.4cm}
{
\begin{align}
	D &= \frac{S_l}{S_0}\frac{S_0}{B}B
	   + 2a\Big(\sum_{i=1}^{f\frac{S_0}{B}}\frac{i}{\frac{S_0}{B}}
	   + f\big(\frac{S_l}{S_0}\frac{S_0}{B}-f\frac{S_0}{B}\big)\Big)B\nonumber\\
	  &+ \cdots\nonumber\\
	  &+ \frac{S_l}{S_{l-1}}\frac{S_{l-1}}{B}2B
	   + 2a\Big(\sum_{i=1}^{f\frac{S_{l-1}}{B}}\frac{i}{\frac{S_{l-1}}{B}}
	   + f\big(\frac{S_l}{S_{l-1}}\frac{S_{l-1}}{B}-f\frac{S_{l-1}}{B}\big)\Big)B
	\label{eq:vatperfile0}
\end{align}
}

By performing calculations (see Section~\ref{app:vatperfile}), we get: 

{
\begin{align}
	D &= S_l\Big(2l-1 + \frac{aflB}{S_l} + 2afl - af\big(\frac{1-\frac{1}{f^l}}{1-\frac{1}{f}}\big)\Big)
\label{eq:vatperfile1}
\end{align}
}

Figure~\ref{fig:VATapprox}(a) includes this extension as SST-VAT and
shows that it tracks RocksDB behavior even closer compared to VAT.

Figure~\ref{fig:VATbase}(b) shows how VAT captures the effect of
designs that exhibit a reduced value for $a$ (reduced amplification)
resulting in lower $r$ (reduced device throughput) as well.  Kreon
uses a value log and small-size SSTs with an index. Small SSTs allow
for a reduced $a.$ At the same time, Kreon exhibits a lower $r$ due to
the randomness introduced from small SSTs. With modern storage
devices, experiments showed that $r$ for Kreon is around 0.91, which
is close to the optimal value of 1. Figure~\ref{fig:VATapprox}(b)
shows that the measured values from Kreon are close to the cost
calculated by VAT. We note that LSM~\cite{lsm} cost analysis does not
describe designs similar to Kreon, therefore, we do not include a
curve from LSM-type analysis.

BlobDB tries to reduce amplification by also using a value log and
merging only metadata (keys and pointers) during compaction. VAT
modeling for leveling with a value log successfully captures this
behavior, as shown in Figure~\ref{fig:VATapprox}(c). Amplification is
reduced significantly because of value separation and the use of small
keys in our workload (default for YCSB).

BlobDB exhibits a value of 0.8 for $a,$ compared to 0.68 in RocksDB
(Table~\ref{tab:atable}). Although both systems use leveling, the use
of the value log in BlobDB results in more keys per SST compared to
RocksDB where SSTs contain both keys and values. As a result, one SST
in BlobDB, typically overlaps with more SSTs of the next level,
resulting in a higher value for $a.$ At the same time, both systems
achieve the same device throughput, as they use similar size SSTs.

In Figure~\ref{fig:VATapprox}(d) VAT models tiering and PebblesDB that
uses a form of tiering. VAT indicates that a larger growth factor
should result in in less amplification.  However, we note that
PebblesDB does not decrease amplification with the same rate as VAT
does because it is not a pure tiering system. The reason for this is
that PebblesDB tries to improve read behavior as follows. To reduce
the number of SSTs that need to be examined during a read operation,
it maintains overlapping SSTs only within guards which results in
higher amplification during compactions. Therefore, although VAT
captures the cost of ``pure'' tiering, PebblesDB exhibits higher cost
in the insert-path. Both exhibit a reducing trend as growth factor
increases, as is expected for amplification in systems that use
tiering.

\section{Tradeoff analysis}
\label{sec:proj}

In this section, we use VAT to examine tradeoffs across different
design points and make additional observations.

\paragraph{The effects of randomness:} Figure~\ref{fig:VATbase}(a)
shows the curves of base VAT for different values of $a$ while
maintaining optimal throughput with $r=1$.  As $a$ decreases,
indicating systems that make use of randomness to reduce
amplification, the optimal number of levels that minimize
amplification decreases as well. Minimum amplification drops from
about 25x to less than 10x, when $a$ decreases from 1.0 to 0.1.
Therefore, techniques that make use of randomness to reduce $a$ can
lead to increased KV-store efficiency. Secondarily, we see for all
values of $a,$ an inappropriate number of levels (small or large)
leads to very high amplification, exceeding 50x for small numbers of
levels, which implies large growth factors.

\paragraph{Optimal growth factor:} The analysis in
Appendix~\ref{app:vatlf} shows that the growth factor between any two
consecutive levels must either be the same or converge to the same
value. Also, we note that the optimal growth factor is constant
regardless of the dataset size.  Therefore, as data grows, both VAT
and LSM dictate that to minimize amplification we have to increase the
number of levels, as opposed to increasing their relative size.
Figure~\ref{fig:LSMVAT}(a) and Figure~\ref{fig:VATbase}(a) plot
amplification as a function of the number of levels for different
values of $a$ and $C$. In both cases, the part of each curve to the
left of the optimal number of levels is more ``steep'' than the part
of the curve to the right. This means that to store the same amount of
data in multi-level designs that perform leveling, and if it is not
possible to use the optimal number of levels, e.g. because of other
considerations, it is preferable in terms of amplification to err
towards using a larger number of levels (and lower $f$) than the
opposite.

\paragraph{Space amplification matters as well:} Many systems choose
larger than optimal growth factors to improve space efficiency. If we
assume that intermediate levels are usually full with updates that
will be garbage collected during compactions, then intermediate levels
incur space amplification, which increases device cost. Space
amplification (excluding the size of user data) can be roughly
calculated as
$\frac{S_0+\ldots+S_{l-1}}{S_l}=\frac{1}{f}+\ldots+\frac{1}{f^l}.$
Using $f=4$ results in space amplification larger than 25\%, which
might be considered excessive, especially for expensive storage
devices, such as SSDs and NVMe. If we use $f=10$ then space
amplification drops to about 10\%. VAT shows that for $a=1$,
$\frac{T}{Topt}=32$ for $f=10$ and $\frac{T}{Topt}=23.91$ for $4$, so
increasing $f$ from $4$ (close to optimal) to $10$ makes amplification
in the insert-path worse by about 1.33x, which is an acceptable cost
for reducing space amplification by 2.5x (from 25\% down to 10\%).
Therefore, VAT allows system designers and users to tune the system
design or configuration.

\paragraph{Single tier for future fast storage devices:} As technology
improves, devices will be able to achieve maximum throughput for even
smaller block sizes, e.g. about 256 bytes for recent NVM
devices~\cite{swanson}). This will allow KV stores to use even smaller
SSTs, further decreasing the value of merge amplification $a.$ If we
assume that $a$ can become 0, this would result in KV stores with a
single level, as indicated by Equation~\ref{eq:xFInewdevices}:

\begin{align}
\frac{T}{T_{opt}} = \frac{2l-1 - al + afl}{r}\xRightarrow[\text{$r=1$}]{\text{$a=0$}}
\frac{T}{T_{opt}} = 2l-1\label{eq:xFInewdevices}
\end{align}

However, we should note that having a value of $a=0$ may not be
possible for arbitrary key sizes, unless devices become truly byte
addressable and are used as such.

\begin{figure}
\centering
\subfigure[$f=10$,~$C=1000$]{\includegraphics[width=0.24\textwidth]{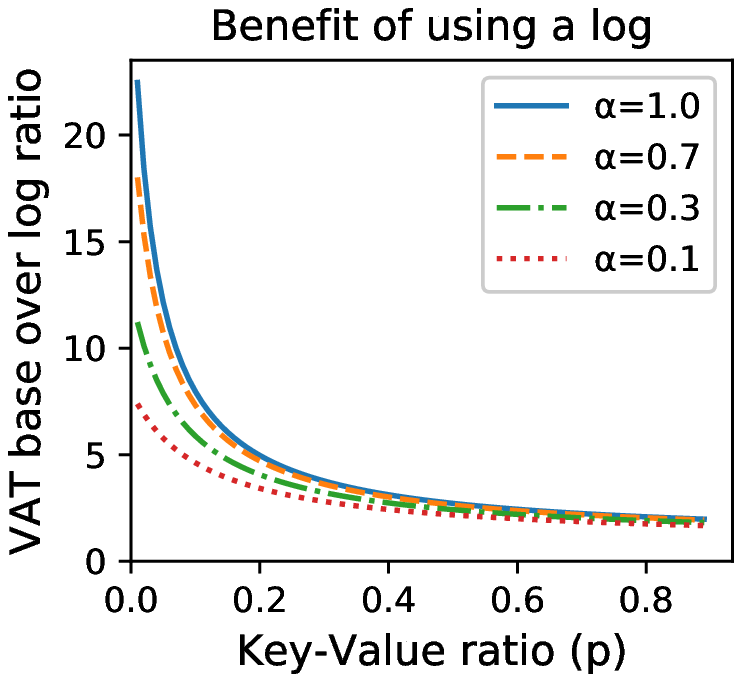}}~
\subfigure[$p=1\%$,~$r=1$, $C=1000$]{\includegraphics[width=0.24\textwidth]{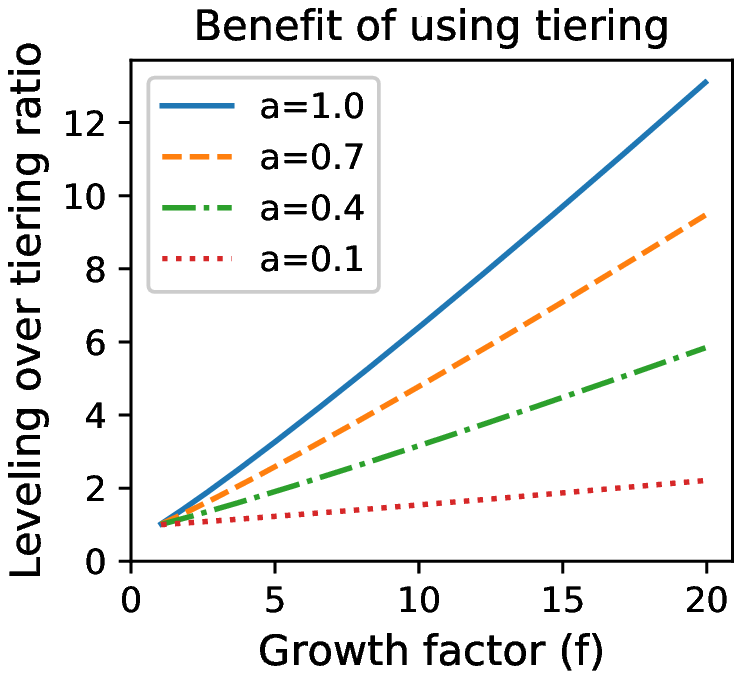}}
\caption{ (a) Amplification benefit of using a value log compared to
  placing the values in place.
(b) Amplification benefit of tiering compared to leveling compaction.
}
\label{fig:tieringlog}
\end{figure}

\paragraph{Key-Value separation on future devices:}
Figure~\ref{fig:tieringlog}(a) shows that using a value log brings
significant benefits when the key-value size ratio is small, about
$p=1\%.$ In addition, the value of $a$ affects the benefits of using a
value log. VAT shows in Figure~\ref{fig:tieringlog}(a) that for
values $a \ge 0.3$ the benefit is more than 10x. However, as device
technology evolves, it allows for optimal device throughput with
smaller unit size, and $a$ can reduce, e.g. to $0.25$ for certain
configurations in Kreon (Table~\ref{tab:atable}). Using
Equation~\ref{eq:futurelog} VAT shows that the benefit of using a log
as $a$ approaches $0$ is given by:

\begin{align}
	\frac{\frac{2l-1-al+afl}{r}}{\frac{p(2l-1-al+afl)+p+1}{r\cdot(p+1)}}
	\xrightarrow[\text{$l=1$}]{\text{$a=0$}}
	\frac{p+1}{2p+1}
	\label{eq:futurelog}
\end{align}

So for any key-value size ratio Equation~\ref{eq:futurelog} shows that
KV separation is actually worse that placing the values in-place with
keys, especially when introducing the extra cost of garbage collection
(see Section~\ref{sec:discuss}).

\label{sub:ss8}

\begin{figure}
\centering
\subfigure[$p=1\%$,~$r=1$, $C=1000$]{\includegraphics[width=0.24\textwidth]{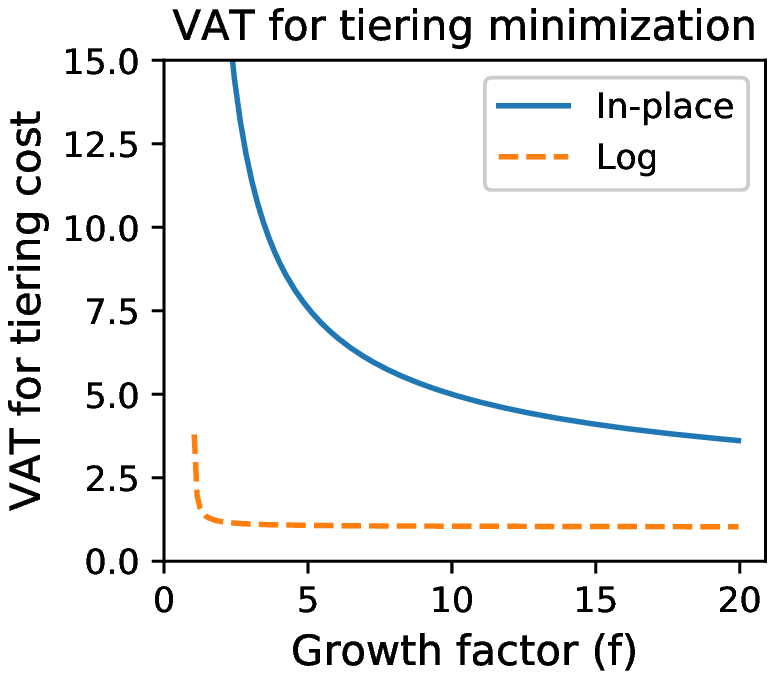}}~
\subfigure[$r=1$ ~$C=1000$]{\includegraphics[width=0.24\textwidth]{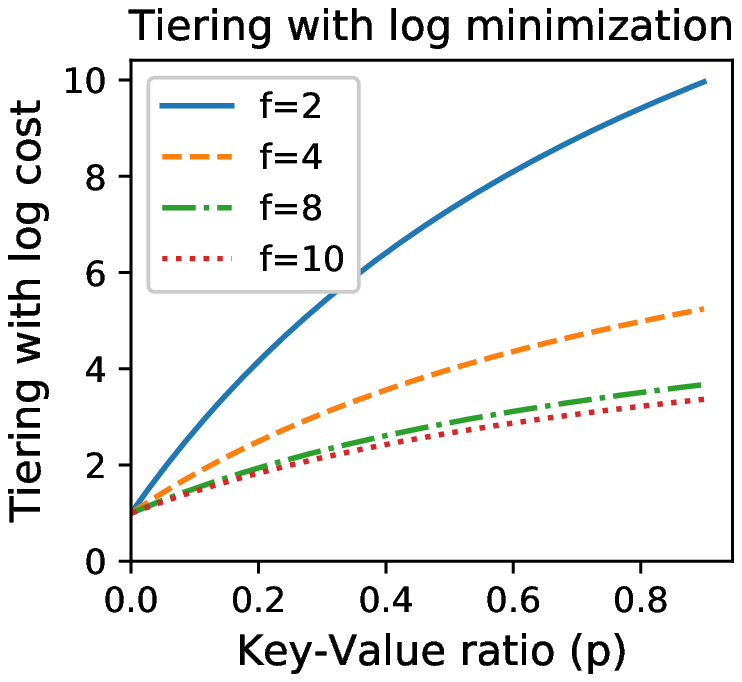}}
\caption{
	(a) VAT for tiering.
	(b) VAT for tiering with values in log.
}
\label{fig:tiering}
\end{figure}

\paragraph{Does tiering still make sense with future devices?} In
Figure~\ref{fig:tiering}(a), VAT shows that when using tiering with
in-place values, one should probably use a growth factor $f$ of around
10, which is the point of diminishing returns for amplification.
After $f=10,$ amplification reduces less, while read operations
continue to become slower. VAT also expresses the fact that with
tiering, throughput does not depend on merge amplification. Thus,
tiering cannot benefit from improved device technology and the ability
to reduce $a.$ Furthermore, Figure~\ref{fig:tieringlog}(b) shows that
the benefit of tiering reduces as merge amplification decreases: As
compactions become cheaper, i.e. $a$ reduces, the benefit of tiering
becomes smaller.

Tiering techniques so far have placed values along with
keys~\cite{stepmerge} without using a log. In
Figure~\ref{fig:tiering}(b) VAT shows that tiering with a log can be a
valid approach for small values of $f$, which do not reduce read
performance significantly and for small $p$ ratios, up to $10\%$.
However, in this case, Figure~\ref{fig:tieringlog}(a) shows that the
amplification benefit is also significant for leveling with a value
log, about 10x. In addition, leveling with a value log has better scan
performance compared to tiering with a value log. Therefore, for small
$p$ ratios, it is preferable to use a value log with leveling.

\section{Related Work}
\label{sec:related}

In this section we first present a taxonomy of various KV-store
designs and then relate this space to VAT.

\paragraph{Taxonomy:} Table~\ref{table:taxonomy} provides a high-level
taxonomy of existing systems that to some extend have tried to take
advantage of device properties and improve performance, similar to
what VAT models. In this taxonomy, we use three dimensions:

\vspace*{.1cm}
\noindent
\underline{{\it Size and placement of SSTs:}}
The SST size used to organize data within each level and their
placement on the device is typically large.  \emph{Large SSTs}
guarantee maximum device throughput, eliminating the effects of the
I/O pattern (sequential or random) and metadata I/Os, as metadata is
small and fits in memory. \emph{Small and sequentially} placed SSTs on
the device~\cite{sears2012blsm} can achieve the same goal of maximum
device throughput. This results in high I/O amplification but improves
read performance at the expense of maintaining more metadata.
Emerging device technologies allow using \emph{small SSTs with random
placement} which introduces randomness but has the potential to reduce
I/O amplification~\cite{kreon}.  This approach is suitable for devices
where random I/O throughput degrades gracefully compared to sequential
I/O throughput.

\vspace*{.1cm}
\noindent
\underline{{\it Logical level organization:}}
Keys in each level are logically organized either \emph{fully} or
\emph{partially}. Full organization keeps the key space in fully
sorted, non-overlapping SSTs. Full organization is usually done with
leveling compaction~\cite{lsm, rocksdb, locs, triad, dostoevsky,
mutant, cLSM, atlas, wisckey, hashkv}. However, B+-tree indexes have
also been used to either optimize reads and scans~\cite{sears2012blsm}
or reduce amplification~\cite{kreon}. Partial organization maintains
the key space in overlapping units, e.g. in the form of tiering
compaction~\cite{lsmtrie,Dayan:2017:MON:3035918.3064054,
Raju:2017:PBK:3132747.3132765, sifrdb,novelsm} which reduces merge
amplification at the cost of reduced read and scan performance.

\vspace*{.1cm}
\noindent
\underline{{\it Value location:}}
Finally, values can be placed either \emph{in-place} with keys or in a
separate \emph{value log}. Typically, values are stored in-place
because this results in optimal scan behavior at the expense of
increasing amplification due to value movement during merge
operations. Previous work has proposed techniques~\cite{hashkv, atlas,
wisckey, tucana, kreon} that store values in a log, reducing
amplification significantly, relying on modern devices to alleviate
the impact on scan performance.

VAT is able to describe the design points within this taxonomy and
quantify tradeoffs across these designs, as device technology
improves. 

\begin{table}[]
\centering
\footnotesize{
\begin{tabular}{l|l|l|l}
                                                 &SST size, & Organi-      & Value    \\ 
                                                 &placement & zation       & placement\\ \hline
LSM\cite{lsm}, RocksDB\cite{rocksdb},            & Large    & Full         & In-place \\
Locs\cite{locs}, Dostoevsky\cite{dostoevsky},    &          &              &          \\
Triad\cite{triad}, Mutant\cite{mutant},          &          &              &          \\
bLSM\cite{sears2012blsm}, cLSM\cite{cLSM}        &          &              &          \\ \hline
Atlas\cite{atlas}, WiscKey\cite{wisckey},        & Large    & Full         & Log      \\
HashKV\cite{hashkv}                              &          &              &          \\ \hline
LSM-trie\cite{lsmtrie},
Monkey\cite{Dayan:2017:MON:3035918.3064054},     & Large    & Partial      & In-place \\
SifrDB\cite{sifrdb}, Novelsm\cite{novelsm}       &          &              &          \\
PebblesDB\cite{Raju:2017:PBK:3132747.3132765}    &          &              &          \\ \hline
Kreon\cite{kreon}				 & Small    & Small        & Log      \\ \hline
$B^e$-tree\cite{betree2015}                      & Small    & Full         & In-place \\ \hline

\end{tabular}
}
\caption{Taxonomy of the main approaches to design KV stores in three
	dimensions.}
\label{table:taxonomy}
\end{table}

\paragraph{LSM-Tree cost analysis:}
\label{sub:lsm}
The LSM-Tree analysis~\cite{lsm} quantifies the tradeoff between
device throughput and amplification and shows that asymptotically it
is better to increase device throughput at the cost of high
amplification for HDDs. VAT generalizes this analysis to a much
broader collection of KV-store design techniques, which includes
those that are dictated by modern device technology.

\begin{figure}[t]
\centering
\subfigure[LSM]{\includegraphics[width=.24\textwidth]{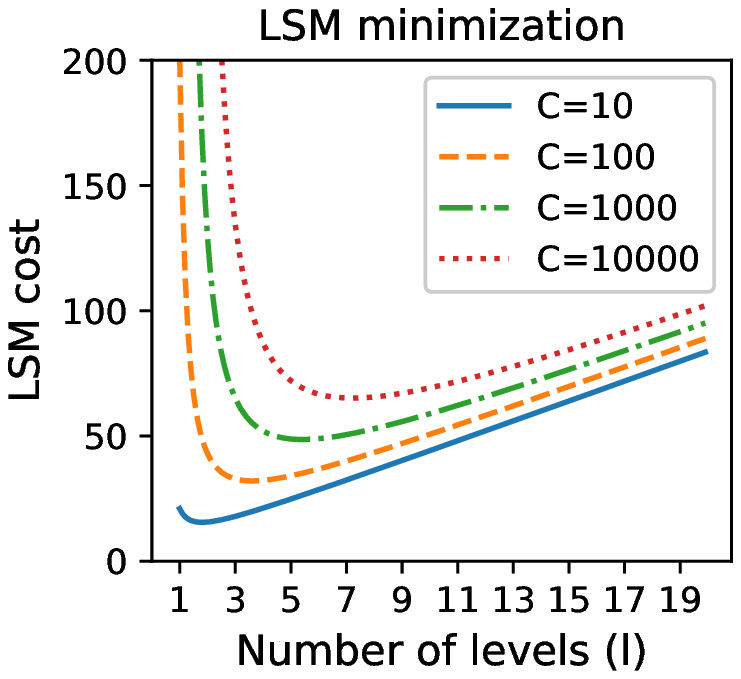}}~
\subfigure[VAT, $a=r=1$]{\includegraphics[width=.24\textwidth]{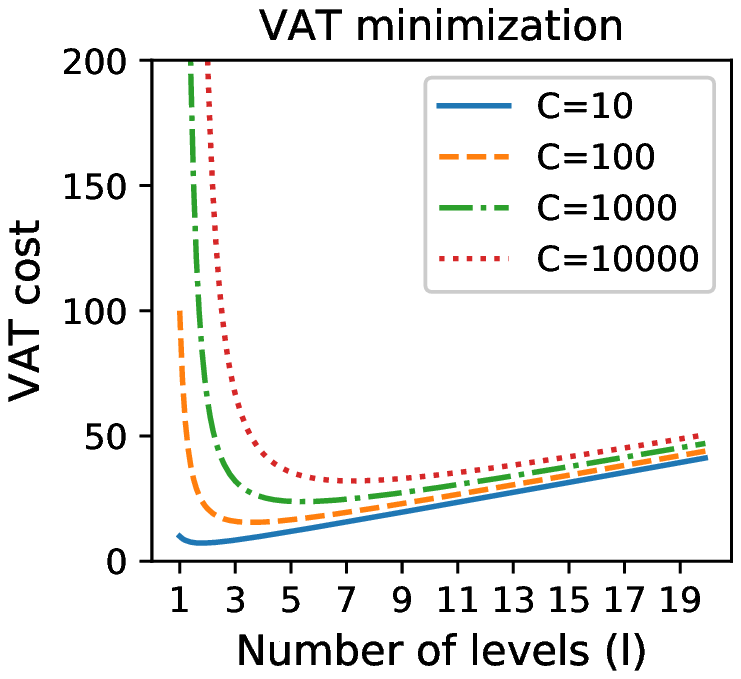}}
\caption{
	(a) Amplification vs. number of levels as a result of
	minimizing the LSM cost for various
	$C=\frac{Workload~size=S_l}{DRAM~size=S_0}$ ratios.
	(b) Amplification vs. number of levels as a result of
	minimizing VAT cost for various $C$ ratios.
	}
\label{fig:LSMVAT}
\end{figure}

Merge and level amplification are competing quantities: As we see in
Figure~\ref{fig:LSMVAT}(a), for a given amount of data, if we reduce
the number of levels $l$, we have to increase the growth factor $f$
between two consecutive levels.  Similarly, reducing the growth factor
$f$, results in increasing the number of levels $l$. Therefore, in a
proper LSM design there is a need to balance $l$ and $f$ to minimize
the total I/O amplification. The LSM-Tree analysis~\cite{lsm} solves
this minimization problem in two cases: (1) when the size ratio of the
entire dataset to main memory is constant and (2) when the sum of the
capacities of all levels is constant. In Appendix~\ref{app:lsmp}, we
use the equations in~\cite{lsm} to solve the minimization problem
without applying the formula simplification used in~\cite{lsm} that
minimizes the total I/O amplification.

\paragraph{Alternative KV-store cost analyses:} The authors
in~\cite{lsmmodel} propose a model that calculates the cost of write
and read amplification in multi-stage log-structured KV-store designs,
offline without the need to run long experiments. Their model takes
into account redundancy (duplicate keys) and uses this to estimate
overheads based on unique keys. They find that the proposed approach
can accurately model the performance of LevelDB and RocksDB and they
show how the model can be used to improve the insert cost in both
systems. VAT shares the goal of modeling write amplification in
multi-stage KV stores (see Figure~\ref{fig:LSMVAT}(b)), however, aims
to capture the behavior of the underlying device technology (parameter
$r$) and also the impact of merge amplification (parameter $a$) of
several design factors. VAT does not take into account redundancy in
the key population but rather aims to capture asymptotic behavior of
different designs and approaches. Our results show that VAT can
describe several existing techniques and design tradeoffs.

MonkeyDB~\cite{Dayan:2017:MON:3035918.3064054} and
Wacky~\cite{dayan2019log} propose and use models to explore tradeoffs
in the design of multi-level KV stores. They focus on optimizing the
use of memory for different aspects of KV stores. Wacky uses this
analysis to search analytically a broad space of merge policies and
other parameters with the goal to identify the parameters that best
match a given workload. The model of Wacky takes into account both the
write and read path and aims to optimize the system for insert, read,
and scan operations. Unlike Wacky, VAT focuses on expressing
asymptotically write amplification and takes into account the impact
of device technology. VAT allows us to explore tradeoffs across design
points in the configuration space, including improving device
technology.

\section{Discussion}
\label{sec:discuss}

In this section we make various observations extracted from our
analysis and evaluation process to provide further insight. We believe
that VAT can be generalized further to (a) take into account
additional important resources and (b) express the cost of additional
operations, in addition to insert path amplification. For instance,
VAT can consider read path amplification, space amplification, optimal
unit of transfer (write, read), etc. Next, we discuss some of these in
more detail.

\paragraph{Unit of transfer in compaction:} In Section~\ref{sec:eval}
we briefly discuss that RocksDB performs compactions in a per-file
manner which results in lower tail latency but increased
amplification. Tail latency is an important metric for data serving
systems that significantly affects the perceived quality of the
system. In KV stores, it is desirable that the system does not block
for long periods when reorganizing data, especially given the volume
of data stored in lower levels.  VAT in Section~\ref{sec:vat},
expresses compactions in a ``stepped'' approach, where each compaction
merges the entire $level_i$ in $level_{i+1}.$ This approach, although
simpler to express analytically, it describes a system where a single
compaction might require that several hundreds of GBs are moved on the
device. This process will take a significant amount of time to
complete and consequently upper levels will be blocked, waiting for
space to be freed. When this effect is propagated to the first levels,
the application will observe long stalls, heavily affecting the
latency of the system. This is the reason why RocksDB merges only a
few SSTs per compaction~\cite{rocksdb} and bLSM~\cite{sears2012blsm}
has proposed a more advanced scheduling technique (gear scheduling).
Each compaction operation is cheaper, which results in much better
latency times and makes the system more responsive. This design
though, comes at the cost of higher amplification: By moving only a
few SSTs per compaction, results in levels that are closer to full at
most times, which makes compactions cumulatively more expensive. An
interesting analysis for future work would be to examine the optimal
number of SSTs to merge in each compaction that brings both
amplification and tail latency down to an acceptable limit.

\paragraph{Garbage collection in the value log:} In
Section~\ref{sec:proj}, VAT shows that key-value separation has a
significant benefit in amplification (up to 10x), especially for small
key-value size ratios of about less than $10\%$ (see
Figure~\ref{fig:tieringlog}(a)). Apart from the fact that there are
cases depending on the key value size that performance of scan
operations degrades significantly, using a value log also incurs space
amplification in workloads with updates. This brings the need for a
garbage collection process, that will periodically examine regions of
the value log and free up updated values to reclaim space. Although
garbage collection in logs is a well studied
problem~\cite{lfs,nova,lfscleaning}, all
solutions~\cite{wisckey,hashkv} incur complexity and cost. For KV
stores, this process incurs several read operations on the multi-level
structure, which is very expensive as multilevel KV stores are not
optimized for this purpose.  VAT suggests that alternative solutions
needs to be explored, using properties specific to KV stores.

In addition, the existence of a log has further implications. If we
use a log, then each SST of fixed size stores more keys. This means
that during compaction each SST will overlap with more SSTs of the
next level with higher probability. Therefore, the existence of a log
tends to result in a higher value for $a,$ with respect to the key
merge operations, see column 4 of Table~\ref{tab:atable}. 

\paragraph{Analytic approximation of $a$:} It would be desirable to
provide an analytic approximation of merge amplification $a$ and
present it as a function of the SST size, the data organization
technique, and the input key distribution. As a first step, it could
be a function that takes as parameters the key distribution and the
SST (bucket) size and provide a probability distribution of how many
buckets will be touched in each compaction. This can allow VAT to
estimate parameter $a$ without the need of actual measurements, at
least for certain input distributions.

\section{Conclusions}
\label{sec:conclusions}

In this paper, we present VAT, an asymptotic analysis that calculates
data amplification in the insert-path for a number of configuration
points in the KV-store design space. We evaluate VAT using RocksDB,
Kreon, BlobDB, and PebblesDB. We show how various design approaches
behave in the insert-path by quantifying their benefits and tradeoffs.
VAT offers significant intuition about the associated tradeoffs: Using
key-value separation decreases amplification down to 1.2x compared to
in-place values which incurs 20x. Tiering compaction reduces
amplification 5x compared to leveling at the cost of reads/scans
operations and is orthogonal to device technology. Introducing
techniques that take advantage of I/O randomness can reduce
amplification from 20x down to 10x. As device technology improves, VAT
suggests that the role of key-value separation and tiering diminishes
and that KV-store designs on aggressive NVM devices may use a single,
in-memory level, to minimize amplification, which is the same concept
with indexing techniques for DRAM. We believe that VAT is useful for
examining tradeoffs and eventually designing KV stores that
dynamically adapt to device properties and increase write performance
by reducing I/O overhead.
%-------------------------------------------------------------------------------
\bibliographystyle{plain}
\bibliography{paper}
%-------------------------------------------------------------------------------
%%\newpage
\appendix
\section{Detailed VAT derivations}
\label{app:vat}

\subsection{Data amplification in base VAT}
\label{app:vatdata} 

Starting from Equation~\ref{eq:vatamp} we can derive
Equation~\ref{eq:vatdata} that expresses the amount of data during
merging in VAT as follows:

\begin{align*}
Eq.~\ref{eq:vatamp} \Rightarrow
%%	D &= \frac{S_l}{S_0}(S_0)  + 2a\sum_{i=1}^{S_l/S_0}((i-1)\bmod{f})\cdot S_0\\
%%	  &+ \frac{S_l}{S_1}(2S_1) + 2a\sum_{i=1}^{S_l/S_1}((i-1)\bmod{f})\cdot S_1\\
%%	  &+ \ldots\\
%%	  &+ \frac{S_l}{S_{l-1}}(2S_{l-1}) + 2a\sum_{i=1}^{S_l/S_{l-1}}((i-1)\bmod{f})\cdot S_{l-1}\Rightarrow\\
	D &= (2l-1)S_l + 2aS_{l-1}\sum_{j=1}^{f^1}(j-1)\bmod{f} \\
	  &+ \ldots + 2aS_0\sum_{j=1}^{f^l}(j-1)\bmod{f}\Rightarrow\\
	D &= (2l-1)S_l + 2aS_{l-1}\Big(\frac{f^1}{f}\cdot \frac{(f-1)(f-1+1)}{2}\Big)\\
	  &+ \ldots + 2aS_0\Big(\frac{f^l}{f}\cdot \frac{(f-1)(f-1+1)}{2}\Big)\Rightarrow\\
	D &= (2l-1)S_l + aS_{l-1}\Big(f^1\cdot (f-1)\Big)\\
	  &+ \ldots + aS_0\Big(f^l\cdot (f-1)\Big)\Rightarrow\\
	D &= S_l(2l-1 - al + afl)
\end{align*}

Then, we can use this equation and $T,T_{opt}$
(Equations~\ref{eq:T and Topt}) to calculate the ratio
$\frac{T}{T_{opt}}$ of Equation~\ref{eq:vat}:

\begin{align*}
	T &= \frac{D}{r\cdot R_{opt}} = \frac{S_l(2l-1 - al + afl)}{r\cdot R_{opt}} = \\
          &= \frac{T_{opt}(2l-1 - al + afl)}{r}\Rightarrow\nonumber\\ 
        \frac{T}{T_{opt}}
          &= \frac{2l-1 - al + afl}{r} 
\end{align*}

\subsection{Data amplification in VAT for key-value separation}
\label{app:vatlog} 

Starting from Equation~\ref{eq:vatlogamp} we can derive
Equation~\ref{eq:vatlogdata} that expresses the amount of data amplification during
merging in VAT as follows:

\begin{align*}
%%D &= \frac{K_l}{K_0}(2K_0)  + 2a\sum_{i=1}^{\frac{K_l}{K_0}}((i-1)\bmod{f}) K_0 + \ldots\\
%%  &+ \frac{K_l}{K_{l-1}}2K_{l-1} + 2a\sum_{i=1}^{\frac{K_l}{K_{l-1}}}((i-1)\bmod{f}) K_{l-1} + S_l\Rightarrow \\
D &= \frac{K_l}{K_0}(K_0)  + 2a\sum_{j=1}^{\frac{K_l}{K_0}}((j-1)\bmod{f})\cdot K_0 + \ldots\\
  &+ \frac{K_l}{K_{l-1}}2K_{l-1} +
  2a\sum_{j=1}^{\frac{K_l}{K_{l-1}}}((j-1)\bmod{f})\cdot K_{l-1} \\
  &+ S_l \Rightarrow\\
  D &= K_l\Big(2l-1 - al + afl\Big) + S_l\\
\end{align*}

Then, we can calculate $\frac{T}{T_{opt}}$ as follows:

\begin{align*}
	T &= \frac{D}{r\cdot R_{opt}} = \frac{K_l(2l-1 - al + afl) + S_l}{r\cdot R_{opt}} = \\
	  &= \frac{S_l(\frac{K_l}{S_l}(2l-1 - al + afl) + 1)}{r\cdot R_{opt}}\Rightarrow\nonumber\\ 
%%          &= \frac{T_{opt}(\frac{K_l}{S_l}(2l-1 - al + afl) + 1}{r}\Rightarrow\nonumber\\ 
\end{align*}
\begin{align*}
        \frac{T}{T_{opt}}
          &= \frac{(\frac{1}{1+\frac{1}{p}})(2l-1 - al + afl) + 1}{r} = \\
%%          &= \frac{(\frac{p}{p+1})(2l-1 - al + afl) + 1}{r} = \\
          &= \frac{p(2l-1 - al + afl) + p + 1}{r\cdot(p+1)}
\end{align*}

\subsection{Optimal growth factor and number of levels in VAT
            under constant $C=\frac{S_l}{S_0}$}
\label{app:vatlf}

\comment{
Optimal growth factor and number of levels in VAT have the same values
as for LSM and the derivation is similar, although the specifics
differ. \comment{We omit this section for space reasons\footnote{We can
  include this in the final version of the paper if space permits.}.}
  }

Starting from our calculation for data amplification for VAT
(Equation~\ref{eq:vatamp}) and assuming that the growth factor $f_i$
can be different at each level, we get:

\begin{align}
Eq.~\ref{eq:vatamp} & \Rightarrow\nonumber\\ 
%%D &= \frac{S_l}{S_0}(S_0)  + 2a\sum_{i=1}^{S_l/S_0}((i-1)\bmod{f_1})\cdot S_0\\
%%  &+ \frac{S_l}{S_1}(2S_1) + 2a\sum_{i=1}^{S_l/S_1}((i-1)\bmod{f_2})\cdot S_1\\
%%  &+ \ldots\\
%%  &+ \frac{S_l}{S_{l-1}}(2S_{l-1}) + 2a\sum_{i=1}^{S_l/S_{l-1}}((i-1)\bmod{f_l})\cdot S_{l-1}\Rightarrow\\
D &= \frac{S_l}{S_0}(S_0)  + 2a\sum_{j=1}^{C}((j-1)\bmod{f_1})\cdot S_0\nonumber\\
  &+ \ldots\nonumber\\
  &+ \frac{S_l}{S_{l-1}}(2S_{l-1}) + 2a\sum_{j=1}^{\frac{C}{f_1\cdots f_{l-1}}}((j-1)\bmod{f_l})\cdot S_{l-1}\Rightarrow\nonumber \\
%%D &= (2l-1)S_l + 2aS_{l-1}\sum_{i=1}^{\frac{C}{f_1\cdots f_{l-1}}}(i-1)\bmod{f_l}\\
%%  &+ \ldots\\
%%  &+ 2aS_0\sum_{i=1}^{C}(i-1)\bmod{f_1}\Rightarrow\\
%%D &= (2l-1)S_l + 2aS_{l-1}\Big(\frac{\frac{C}{f_1\cdots f_{l-1}}}{f_l}\cdot \frac{(f_l-1)(f_l-1+1)}{2}\Big)\\
%%  &+ \ldots\\
%%  &+ 2aS_0\Big(\frac{C}{f_1}\cdot \frac{(f_1-1)(f_1-1+1)}{2}\Big)\Rightarrow\\
%%D &= (2l-1)S_l + 2aS_{l-1}\Big(\frac{\frac{C}{f_1\cdot f_2\cdots f_{l-1}}}{f_l}\cdot \frac{(f_l-1)f_l}{2}\Big)\\
%%  &+ \ldots\\
%%  &+ 2aS_0\Big(\frac{C}{f_1}\cdot \frac{(f_1-1)f_1}{2}\Big)\Rightarrow\\
%%D &= (2l-1)S_l + aS_{l-1}\Big(\frac{C\cdot (f_l-1)}{f_1\cdot f_2\cdots f_{l-1}}\Big)\\
%%  &+ \ldots\\
%%  &+ aS_0\Big(\frac{C\cdot (f_1-1)}{1}\Big)\Rightarrow\\
%%D &= (2l-1)S_l + a\cdot f_1f_2\cdots f_{l-1}\cdot S_0\Big(\frac{C\cdot (f_l-1)}{f_1\cdot f_2\cdots f_{l-1}}\Big)\\
%%  &+ \ldots\\
%%  &+ aS_0\Big(\frac{C\cdot (f_1-1)}{1}\Big)\Rightarrow\\
D %%&= (2l-1)S_l + aCS_0(f_l-1) + \ldots + aCS_0(f_1-1)\Rightarrow\nonumber\\
  %%&= (2l-1)S_l + aCS_0(f_1+f_2+\ldots+f_l-l)\nonumber\\
  %%&= (2l-1)CS_0 + aCS_0(f_1+f_2+\ldots+f_l-l)\nonumber\\
  &= CS_0\Big(2l-1 + a(f_1+f_2+\ldots+f_l-l)\Big)\label{eq:Dbase}
\end{align}

Then we minimize $\frac{T}{T_{opt}}$:

\begin{align*}
    \min_{\prod_{i=1}^{l}{f_i}=\,C}\frac{T}{T_{opt}} 
        = \min_{\prod_{i=1}^{l}{f_i}=\,C}\frac{D}{r\cdot R_{opt} \cdot T_{opt}} =\\  
        \min_{\prod_{i=1}^{l}{f_i}=\,C}\frac{D}{r\cdot S_l} = 
        \min_{\prod_{i=1}^{l}{f_i}=\,C}\frac{2l-1 + a(\sum_{i=1}^{l}f_i - l)}{r}
\end{align*}

Similar to the LSM analysis, we consider the number of levels to be
constant. Unlike the LSM analysis, the minimization problem also
depends on parameters $a$ and $r$ which model the device technology.
Parameters $a$ and $r$ are constants given a device technology.
Although they affect the value of the minimization point and thus, the
optimal growth factor, the analysis is similar to
Appendix~\ref{app:lsm}. \comment{In Figure~\ref{fig:VATbase_dev} VAT shows that
the optimal growth factor decreases as the device technology
improves. \note{figure number is missing.}}

\comment{
amplification a, and throughput r, we can minimize this function
by using the derivative of
$\sum_{i=1}^{l}f_i$, which leads to the same conclusion for the
optimal values of $f,l$ as for the LSM cost model.
}

\subsection{Data amplification in base VAT for per-SST compaction}
\label{app:vatperfile}

Starting from Equation~\ref{eq:vatperfile1} we derive
Equation~\ref{eq:vatperfile2} that expresses data transfers
in a per-SST basis as follows:

\begin{align}
	Eq.~\ref{eq:vatperfile0} & \nonumber\Rightarrow\\ 
	D &= S_l + 2aB\Big(\frac{B}{S_0}\frac{(f\frac{S_0}{B}+1)f\frac{S_0}{B}}{2} + f(\frac{S_l}{B}-f\frac{S_0}{B})\Big)\nonumber\\
	  &+\nonumber\ldots\\
	  &+2S_l + 2aB\Big(\frac{B}{S_{l-1}}\frac{(f\frac{S_{l-1}}{B}+1)f\frac{S_{l-1}}{B}}{2} + f(\frac{S_l}{B}-f\frac{S_{l-1}}{B})\Big)\nonumber\Rightarrow\\
	D &= S_l + 2afB\Big(\frac{\frac{S_1}{B}+1}{2} + \frac{S_l}{B} - \frac{S_1}{B}\Big)\nonumber\\
	  &+\nonumber\ldots\\
	  &+ 2S_l + 2afB\Big(\frac{\frac{S_l}{B}+1}{2} + \frac{S_l}{B} - \frac{S_l}{B}\Big)\nonumber\Rightarrow\\
	D &= S_l(2l-1) + af\Big(S_1+\cdots +S_l\Big) + aflB + 2aflS_l - 2af\Big(S_1+\cdots +S_l\Big)\nonumber\Rightarrow\\
	D &= S_l(2l-1) + aflB + 2aflS_l - af\Big(S_1+\cdots +S_l\Big)\nonumber\Rightarrow\\
	D &= S_l(2l-1) + S_l\frac{aflB}{S_l} + 2aflS_l - afS_l\Big(\frac{1}{f^{l-1}}+\cdots +\frac{1}{f^0}\Big)\nonumber\Rightarrow\\
	D &= S_l\Big(2l-1 + \frac{aflB}{S_l} + 2afl - af\big(\frac{1-\frac{1}{f^l}}{1-\frac{1}{f}}\big)\Big)\label{eq:vatperfile2}
\end{align}

\section{Detailed derivations of LSM}
\label{app:lsm}

In this appendix for completeness we re-iterate the original
LSM-tree~\cite{lsm} analysis for the insert cost by clarifying how
certain quantities in the analysis relate to device throughput. LSM
defines the \textit{total page I/O rate H} as the required rate at
which a system should operate to handle an incoming data rate R. For
instance, if we assume an incoming rate of R and an amplification of A
then $H=R\cdot A$. In the LSM design, HDDs always operate at the
maximum throughput and H can be achieved by using more HDDs. However,
it is still important to minimize amplification and therefore the
number of disks required to handle a desired data rate R. Based on
this line of thought, H can be defined as:

\begin{align}
%%	H &=\frac{R}{S_p}((2f_1+2) + (2f_2+2) + ... + (2f_l+1))\label{eq:H}
	H &=\frac{R}{S_p}((2f_1+2) + ... + (2f_l+1))\label{eq:H}
\end{align}

\begin{itemize}
\itemsep0em 
\item $R$ bytes per second
\item $S_p$ bytes per page
\item $l$ is the number of levels
\item $f$ variables represent size ratios between adjacent
	levels $f_i=\frac{S_i}{S_{i-1}}$
\item $2f_i+l$ represents all I/O on level $l_i$
\item $f_i\frac{R}{S_p}$ to read in pages from $l_i$ for the
	merge from $l_{i-1}$ to $l_i$
\item $(f_i+1)\frac{R}{S_p}$ to write out pages to $l_i$ for the
	same merge
\item $\frac{R}{S_p}$ to read in pages from $l_i$ for the merge
	from $l_i$ to $l_{i+1}$
\end{itemize}

An important note here is that H is based on the assumption that in
every spill from $level_{i}$ to $level_{i+1}$, we read and write the
entire $level_{i+1}$ once. We can re-write $H$ as:

\begin{align}
  H &=\frac{R}{S_p}(2(f_1+f_2+...+f_l) + 2l - 1)\nonumber\\
    &=\frac{R}{S_p}(2\sum_{i=1}^{l}f_i + 2l - 1)\label{eq:lsmH}
\end{align}

To minimize H, the authors in \cite{lsm} observe that the rate of
insertions to all levels is the same at steady state (and $S_p$ is
constant as well), so it suffices to minimize the second factor of
H. Therefore:

\begin{align}
  \min H = \min (2\sum_{i=1}^{l}f_i + 2l - 1) \label{eq:lsmmin}
\end{align}

This minimization problem makes sense and has non-trivial solutions
under some constraint on the amount of data that needs to be
stored. In the original LSM analysis~\cite{lsm} the authors use two
different constraints:
(1) The ratio of DRAM to the dataset size is constant $\frac{S_l}{S_0}=C$: 
\begin{align}
   \min_{\frac{S_l}{S_0}=C}H &= \min_{\frac{S_l}{S_0}=C}(2\sum_{i=1}^{l}f_i + 2l - 1)\label{eq:lsmminpa}
\end{align}
(2) The total size of all levels is constant $S_0+...+S_l=C$:
\begin{align}
   \min_{S_0+...+S_l=C}H &= \min_{S_0+...+S_l=C}(2\sum_{i=1}^{l}f_i + 2l - 1) \label{eq:lsmminpb}
\end{align}

The practical meaning of the first constrain is that the size of the
workload $S_l$ is a function of DRAM size. This direct correlation,
simplifies the minimization problem further down in the analysis. On
the other hand, the second constrain is more relaxed, in a sense that
it only bounds the storage capacity the system has, which better fits
real scenarios but results in a harder minimization problem. We also
note that in the first constrain $C$ is only a scalar whereas in the
second constrain $C$ is measured in bytes.

Next, we present the solution of the minimization problem for the
first case. \comment{The conclusions of the second case are the same,
however, we omit due to space limitations the detailed
solution\footnote{We can include this in the final version of the
paper for completeness, if space permits}. As a summary, in the
second case we notice that the optimal growth factor $f_i$ between two
levels $i$ (large) and $i-1$ (small) reduces by a small decrement at
each $i$. For instance, for $l=5$ levels, an example sequence is the
following: $f_1\approx 11.0998, f_2\approx 11.0991, f_3\approx
11.0909, f_4=11, f_5=10$. Essentially, these values have small
differences and for all practical purposes they can be considered to
be the same value $f_5$. Therefore, even with the constant total size
assumption, it turns out that for all practical purposes the growth
factor is constant across levels and we can follow the same process as
in the previous section to calculate the optimal $l,f$.}

\subsection{LSM optimal growth factor and number of levels under constant $C=\frac{S_l}{S_0}$}
\label{app:lsmp} 

In the case of constant $\frac{S_l}{S_0}=C$ and since $S_i = f_i\cdot
S_{i-1}$ we can write the constraint as a product of $f_i$: $S_l =
\prod_{i=1}^{l}{f_i}\cdot S_0$ or $\frac{S_l}{S_0} =
\prod_{i=1}^{l}{f_i}$ or $\prod_{i=1}^{l}{f_i} = C$. Therefore:

\begin{align*}
   \min_{\prod_{i=1}^{l}{f_i}=\,C}H &= \min_{\prod_{i=1}^{l}{f_i}=\,C}(2\sum_{i=1}^{l}f_i + 2l - 1)
\end{align*}

The original LSM analysis~\cite{lsm} argues that based on this
formulation we can conclude that all $f_i$ are the same. We elaborate
this argument, as follows. If we fix the number of levels to any constant value L, meaning that
we examine every possible number of levels in a KV store, then the
minimization problem becomes:

\begin{align*}
   \min_{\prod_{i=1}^{l}{f_i}=\,C}H &= \min_{\prod_{i=1}^{l}{f_i}=\,C}(2\sum_{i=1}^{l}f_i + 2L - 1)
\end{align*}

To solve this minimization problem we can replace the $f_l$ term in
the sum using the constraint. So the minimization problem can be
written as:

\begin{align*}
   \min_{\prod_{i=1}^{l}{f_i}=\,C}\sum_{i=1}^{l}f_i
        &= \min(2\sum_{i=1}^{l-1}{f_i+C\cdot\prod_{i=1}^{l-1}{f_i^{-1}}} + 2L - 1)
\end{align*}

Finding the minimum of a function is equivalent to finding the point
where its derivative is equal to 0. Using partial derivatives for each
of the free variables $f_j, j=1,\ldots,l-1$ we get:

\begin{multline*}
(2\sum_{i=1}^{l-1}{f_i+C\cdot\prod_{i=1}^{l-1}{f_i^{-1}}} + 2L - 1)\frac{d}{df_j} =0\Rightarrow\\
(f_1+f_2+\ldots+f_{l-1}+C\cdot\prod_{i=1}^{l-1}{f_i^{-1}} + 2L - 1)\frac{d}{df_j} =0
\end{multline*}

Given that L is a constant, we get:

\begin{align*}
	1-\frac{1}{f_j}\cdot C\cdot\prod_{i=1}^{l-1}{f_i^{-1}} &= 0\Rightarrow
        f_j=C\cdot\prod_{i=1}^{l-1}{f_i^{-1}} 
\end{align*}

Therefore, all $f_j$ must have the same value $f=f_1=...=f_l$ and leads to:
leads to:

\begin{align}
	C=\,f_1\cdot f_2\cdot\ldots\cdot f_l \Rightarrow C=\,f^l \Rightarrow f=\,\sqrt[l]{C} \label{eq:lsmf}
\end{align}

If we replace this value $f$ in Equation~\ref{eq:lsmminpa}, we get:

\begin{align*}
   \min_{\frac{S_l}{S_0}=C}H &= \min_{\frac{S_l}{S_0}=C}(2lf + 2l - 1)
                             = \min_{\frac{S_l}{S_0}=C}(2l \cdot \sqrt[l]{C} + 2l - 1)
\end{align*}

To minimize this function we solve for the point where its derivative
becomes 0:

\begin{multline*}
\frac{d}{dl}(2l \cdot \sqrt[l]{C} + 2l - 1) = 0 \Rightarrow 2\sqrt[l]{C} - \frac{2\sqrt[l]{C}\cdot log_eC}{l} + 2 = 0\\
\Rightarrow \sqrt[l]{C} \Big(\frac{log_eC}{l}-1\Big) = 1\Rightarrow (e^{log_eC})^\frac{1}{l}\Big(\frac{log_eC}{l}-1\Big) = 1\\
\Rightarrow e^\frac{log_eC}{l}\Big(\frac{log_eC}{l}-1\Big) = 1 \xRightarrow[]{\text{$\frac{log_eC}{l}=x$}} e^x(x-1) = 1 \\
\Rightarrow (x-1)e^xe^{-1} = e^{-1} \Rightarrow	(x-1)e^{x-1} = \frac{1}{e} 
\end{multline*}

If we use Lambert's W function~\cite{lambertf}, then by definition $W(xe^x)=x$. Therefore:

\begin{align*}
W((x-1)e^{x-1}) = W(\frac{1}{e}) \Rightarrow x-1 = W(\frac{1}{e}) \xRightarrow[]{\text{$x=\frac{log_eC}{l}$}}\\
\frac{log_eC}{l}-1 = W(\frac{1}{e}) \Rightarrow l = \frac{log_eC}{W(\frac{1}{e})+1}
\end{align*}

We can now calculate the optimal
growth factor $f$ as:

\begin{align}
f &= \sqrt[l]{C} = C^\frac{1}{l} = C^\frac{1}{\frac{log_eC}{W(\frac{1}{e})+1}} = C^\frac{W(\frac{1}{e})+1}{log_eC} \nonumber\\
  &= (C^\frac{1}{log_eC})^{W(\frac{1}{e})+1} = (e^{log_eC\frac{1}{log_eC}})^{W(\frac{1}{e})+1}\nonumber\\ 
  &= e^{W(\frac{1}{e})+1} 
\end{align}

Given that $W(\frac{1}{e})$ is about 0.5~\cite{lambertcalculator}, we
can write $l=\frac{log_eC}{1.5}$ and $f=e^\frac{3}{2}$.

We can derive a somewhat less accurate but simpler value for the
optimal number of levels and growth factor by solving the simplified
minimization problem:

\begin{align*}
     %%\min_{\prod_{i=1}^{l}{f_i}=\,C}H =
     \min_{\prod_{i=1}^{l}{f_i}=\,C}(2lf + 2l -1)
             &\approx \min_{\prod_{i=1}^{l}{f_i}=\,C}(lf)
\end{align*}

Similar to above, using derivatives leads to:

\begin{multline}
	\frac{d}{dl}\Big( l\cdot \sqrt[l]{C}\Big) = 
	\sqrt[l]{C} + l\Big(log_eC\cdot \sqrt[l]{C}\cdot (-\frac{1}{l^2})\Big) \\
	= \sqrt[l]{C}\Big(1 - \frac{log_eC}{l}\Big) = 0\Rightarrow l = log_eC\label{eq:simplel}
\end{multline}

Therefore, the optimal number of levels is $l=log_eC$ and we can
calculate the optimal growth factor as
$f=\sqrt[l]{C}=\sqrt[log_eC]{C}=e$.

\subsection{Optimal growth factor and number of levels under constant total size $C=S_0+...+S_l$}
\label{app:lsms} 

In the second case, we solve for a different assumption. Simplifying
the minimization problem as in the previous case:

\begin{align*}
\min_{S_0+...+S_l=C}H &= \min_{S_0+...+S_l=C}(2\sum_{i=1}^{l}f_i + 2l - 1)
                 \approx \min_{\sum_{i=0}^{l}{S_i=C}} \sum_{i=1}^{l}{f_i}
\end{align*}

If we use the \textit{Lagrange Multipliers,} the problem can be
described as follows:

\begin{align*}
	L(f_1,f_2,\ldots,f_l,\lambda) =
	h(f_1,f_2,\ldots,f_l)-\lambda(g(S_0,S_1,\ldots,S_l)-C)
\end{align*}

\noindent where $h(f_1,\ldots,f_l)$ is the function we want to
minimize, $g(S_0,S_1,\ldots,S_l)$ is the function describing the
constraints and $\lambda$ is called the \textbf{Lagrange Multiplier}.
To find the minimum produced by a certain set of values for $f_1, f_2,
\ldots, f_l$ we proceed as follows:

\begin{align*}
	\nabla L = 0 \Rightarrow
	\left(\frac{\partial L}{\partial f_1}, \frac{\partial L}{\partial
	f_2}, \ldots, \frac{\partial L}{\partial f_l}\right) = 0
\end{align*}

So now we take each dimension equal to 0 to satisfy the equation:

\begin{align*}
	\frac{\partial L}{\partial f_i} = 0, i\in[1,l]
\end{align*}

Taking the partial derivative of $L$ with respect to $f_1$ we get:

\begin{align*}
	\frac{\partial L}{\partial f_1} &= 0\Rightarrow\\
	\frac{\partial h(f_1,f_2,\ldots,f_l)}{\partial f_1} -
	\lambda\cdot
	\frac{\partial g(S_0,S_1,\ldots,S_l)}{\partial f_1} -
	\lambda\cdot \frac{\partial C}{\partial f_1} &= 0 \Rightarrow\\
	1-\lambda\cdot (S_0 + f_2S_0 + \cdots + f_2\cdot f_3\cdots f_l
	\cdot S_0) &= 0 \Rightarrow\\
        S_0 + f_2S_0 + \cdots + f_2\cdot f_3\cdots f_l\cdot S_0 &= \frac{1}{\lambda}
\end{align*}

If we rewrite:

\begin{align*}
        (S_0 + f_2S_0 + \cdots + f_2\cdot f_3\cdots f_l \cdot S_0) = \frac{C-S_0}{f_1} \Rightarrow \\
	\frac{1}{\lambda} = \frac{C-S_0}{f_1}\Rightarrow 
        f_1 = \lambda\cdot (C-S_0)
\end{align*}

Equivalently, we can calculate:

\begin{align*}
	f_1 &= \lambda\cdot (C-S_0)\\
	f_2 &= \lambda\cdot (C-S_0-f_1S_0)\\
	f_3 &= \lambda\cdot (C-S_0-f_1S_0-f_1f_2S_0)\\
	\ldots\\
	f_l &= \lambda\cdot
	(C-S_0-f_1S_0-f_1f_2S_0-\ldots-\prod_{i=1}^{l}f_i S_0)
\end{align*}

If we perform the subtractions, the early $f_i$'s will be constituted
by a large number of terms but in the latter $f_i$'s, most of the
terms will be canceled out. This way:

\begin{align*}
	f_l &= -\lambda\cdot\prod_{i=1}^{l}f_i S_0\Rightarrow\\
	f_{l-1} &= -\lambda\cdot\left(\prod_{i=1}^{l-1}f_i
	S_0+\prod_{i=1}^{l}f_i S_0\right)\Rightarrow\\
	f_{l-2} &= -\lambda\cdot\left(\prod_{i=1}^{l-2}f_i
	S_0+\prod_{i=1}^{l-1}f_i S_0+\prod_{i=1}^{l}f_i
	S_0\right)\Rightarrow\\
	\ldots
\end{align*}

It is easier to start from the higher terms of $f_i$:

\begin{align*}
	\frac{f_{l-1}}{f_l} &=
	\frac{-\lambda\cdot\left(\prod_{i=1}^{l-1}f_i
	S_0+\prod_{i=1}^{l}f_i
	S_0\right)}{-\lambda\cdot\prod_{i=1}^{l}f_i S_0}\Rightarrow\\
	\frac{f_{l-1}}{f_l} &=
	\frac{\prod_{i=1}^{l-1}f_i+\prod_{i=1}^{l}f_i}{\prod_{i=1}^{l}f_i}\Rightarrow\\
	\frac{f_{l-1}}{f_l} &=
	\frac{\prod_{i=1}^{l-1}f_i}{\prod_{i=1}^{l}f_i}+1\Rightarrow\\
	\frac{f_{l-1}}{f_l} &= \frac{1}{f_l}+1\Rightarrow f_{l-1} = 1 + f_l
\end{align*}

We can keep taking fractions to unveil each $f_i$ value:

\begin{align*}
	\frac{f_{l-2}}{f_{l-1}} &=
	\frac{-\lambda\cdot\left(\prod_{i=1}^{l-2}f_i
	S_0+\prod_{i=1}^{l-1}f_i S_0+\prod_{i=1}^{l}f_i
	S_0\right)}{-\lambda\cdot\left(\prod_{i=1}^{l-1}f_i
	S_0+\prod_{i=1}^{l}f_i S_0\right)}\Rightarrow\\
	\frac{f_{l-2}}{f_{l-1}} &=
	\frac{\prod_{i=1}^{l-2}f_i+\prod_{i=1}^{l-1}f_i+\prod_{i=1}^{l}f_i}{\prod_{i=1}^{l-1}f_i+\prod_{i=1}^{l}f_i}\Rightarrow\\
	\frac{f_{l-2}}{f_{l-1}} &=
	\frac{\prod_{i=1}^{l-2}f_i}{\prod_{i=1}^{l-1}f_i+\prod_{i=1}^{l}f_i}+1\Rightarrow\\
	\frac{f_{l-2}}{f_{l-1}} &= \frac{f_1\cdot f_2\cdots f_{l-2}}
	{(f_1\cdots f_{l-2})\cdot f_{l-1}+
	(f_1\cdots f_{l-2})\cdot f_{l-1}\cdot f_l}+1\Rightarrow\\
	\frac{f_{l-2}}{f_{l-1}} &= \frac{1}{f_{l-1}+f_{l-1}\cdot f_l}+1\Rightarrow\\
	\frac{f_{l-2}}{f_{l-1}} &= \frac{1}{f_{l-1}\cdot(1+f_l)}+1\Rightarrow\\
	f_{l-2} &= f_{l-1}+\frac{1}{1+f_l}\Rightarrow\\
	f_{l-2} &= f_{l-1}+\frac{1}{f_{l-1}}\\
\text{Similarly:}\\
	f_{l-3} &= f_{l-2}+\frac{1}{f_{l-1}}\cdot\frac{1}{f_{l-2}}\\
	\ldots \\
	f_1 &= f_2+\frac{1}{f_{l-1}}\cdots\frac{1}{f_2}
\end{align*}

We notice that the optimal growth factor $f_i$ between two levels $i$
(large) and $i-1$ (small) reduces by a small decrement at each
$i$. For instance, for $l=5$ levels, an example sequence is the
following: $f_1\approx 11.0998, f_2\approx 11.0991, f_3\approx
11.0909, f_4=11, f_5=10$. Essentially, these values have small
differences and for all practical purposes they can be considered to
be the same value $f_5$.
Therefore, even with the constant total size assumption, it turns out
that for all practical purposes the growth factor is constant across
levels and we can follow the same process as in the previous section
to calculate the optimal $l,f$.

%-------------------------------------------------------------------------------

%%%%%%%%%%%%%%%%%%%%%%%%%%%%%%%%%%%%%%%%%%%%%%%%%%%%%%%%%%%%%%%%%%%%%%%%%%%%%%%%
\end{document}